\documentclass{aa}
\usepackage{natbib}
\bibpunct{(}{)}{;}{a}{}{,} 
\usepackage{setspace}
\usepackage{graphicx}
\usepackage{epstopdf}
\usepackage{tabularx}
\usepackage{lscape}
\usepackage{longtable}
\usepackage{mathabx}
\usepackage[T1]{fontenc} 
\usepackage{dingbat}
\newcommand{\masyr}{mas\,yr$^{-1}$}

\begin{document}
\title{MASSIVE: A Bayesian analysis of giant planet populations around low-mass stars}

\author{
  J. Lannier \inst{1}
\and  P. Delorme \inst{1}
\and  A.M. Lagrange \inst{1}
\and  S. Borgniet \inst{1}
\and  J. Rameau \inst{2}
\and  J. E. Schlieder \inst{3}
\and  J. Gagn\'e \inst{4,5}
\and  M.A. Bonavita \inst{6}
\and  L. Malo \inst{7}
\and  G. Chauvin \inst{1}
\and  M. Bonnefoy \inst{1}
\and  J.H. Girard \inst{8}
    }
\offprints{J. Lannier,
  \email{justine.lannier@univ-grenoble-alpes.fr}
   }

\institute{Univ. Grenoble Alpes, Institut de Plan\'etologie et d'Astrophysique de Grenoble (IPAG, UMR 5274), F-38000 Grenoble, France
\and Institut de Recherche sur les Exoplan\`etes (iREx), D\'epartement de physique and Observatoire du Mont M\'egantic,
  Universit\'e de Montr\'eal, C.P. 6128, Succursale Centre-Ville,
  Montr\'eal, QC H3C 3J7, Canada
  \and NASA Ames Research Center, Space Science and Astrobiology Division, MS 245-6, Moffett Field, CA, 94035, USA
  \and Department of Terrestrial Magnetism, Carnegie Institution of Washington, Washington, DC 20015, USA
  \and Sagan Fellow
  \and Institute for Astronomy, The University of Edinburgh, Royal Observatory, 
Blackford Hill, Edinburgh, EH9 3HJ, U.K.
\and CFHT Corporation, 65-1238 Mamalahoa Hwy, Kamuela, Hawaii 96743, USA
\and European Southern Observatory, Alonso de Cordova 3107, Vitacura, Santiago, Chile
}
\abstract{Direct imaging has led to the discovery of several giant planet and brown dwarf companions. These imaged companions populate a mass, separation and age domain (mass$>1~M_\mathrm{Jup}$, orbits>5~AU, age<1~Gyr) quite distinct from the one occupied by exoplanets discovered by the radial velocity or transit methods. This distinction could pinpoint that different formation mechanisms are at play.} 
{We aim at investigating correlations between the host star's mass and the presence of wide-orbit giant planets, and at providing new observational constraints on planetary formation models.}
{We observed 58 young and nearby M-type dwarfs in $L^\prime$-band with the VLT/NaCo instrument and used ADI algorithms to optimize the sensitivity to planetary-mass companions and to derive the best detection limits. We estimate the probability of detecting a planet as a function of its mass and physical separation around each target. We conduct a Bayesian analysis to determine the frequency of substellar companions orbiting low-mass stars, using a homogenous sub-sample of 54 stars.} 
{We derive a frequency of $4.4^{+3.2}_{-1.3}\%$ for companions with masses in the range of 2-80~$M_\mathrm{Jup}$, and $2.3^{+2.9}_{-0.7}$\% for planetary mass companions (2-14~$M_\mathrm{Jup}$), at physical separations of 8 to 400~AU for both cases. Comparing our results with a previous survey targeting more massive stars, we find evidence that substellar companions more massive than 1~$M_\mathrm{Jup}$ with a low mass ratio Q with respect to their host star (Q<1\%), are less frequent around low-mass stars. This may represent an observational evidence that the frequency of imaged wide-orbit substellar companions is correlated with stellar mass, corroborating theoretical expectations. On the opposite, we show statistical evidence that intermediate-mass ratio (1\%<Q<5\%) companion with masses >2~$M_\mathrm{Jup}$ might be independent from the mass of the host star. }
{}

\date{}


\maketitle

\section{Introduction}
After more than two decades of exoplanet discoveries \citep[e.g.][]{Wolszczan.1992, Mayor.1995, Udalski.2002}, more than two thousand planets have been identified, mostly using radial velocity (hereafter RV) and transit methods. Both techniques have detected exoplanets with masses ranging from one Earth mass up to the brown dwarf mass regime ($\sim$13-80~$M_\mathrm{Jup}$), orbiting at close separations from their host star (usually within 5~AU). Direct imaging (DI) has provided data for about 60 large separation ($\ge$10~AU) massive planets and brown dwarfs for more than a decade \footnote{http://exoplanet.eu/} (e.g. 51~Eri~b, \citealt{Macintosh.2015}; HD95086b, \citealt{Rameau.2013.HD95}; GJ504b, \citealt{Kuzuhara.2013}; 2MASS0103(AB)b, \citealt{Delorme.2013}; $\beta$~Pictoris b, \citealt{Lagrange.2010}; HR8799bcde, \citealt{Marois.2008}, \citealt{Marois.2010}; 2M1207b, \citealt{Chauvin.2004}). About a third of the imaged substellar companions (SCs, $<80M_\mathrm{Jup}$) have young M-type dwarf host stars (Table~\ref{Mcompanions}).
\\
DI is usually limited to the detection of distant young massive planets because of their more favorable planet-to-host-star contrast and their larger angular separation. Some of the limitations of DI can be mitigated by using thermal imaging (in $L^\prime$-band) where young massive planets emit most of their flux, and by taking advantage of the intrinsically lower flux of low-mass stars, which provides a more favorable star-planet contrast.
\begin{table*}
\small
\begin{center}
\caption{Giant planets and brown dwarfs imaged around self-luminous young M-dwarfs (source: http://exoplanet.eu). Ref: $^{(1)}$\citet{Delorme.2013}, $^{(2)}$\citet{Naud.2014}, $^{(3)}$\citet{Artigau.2015}, $^{(4)}$\citet{Kraus.2014}, $^{(5)}$\citet{Itoh.2005}, $^{(6)}$\citet{Todorov.2010},$^{(7)}$\citet{Luhman.2006}, $^{(8)}$\citet{Chauvin.2004}, $^{(9)}$\citet{Gauza.2015}, $^{(10)}$\citet{Burgasser.2010}, $^{(11)}$\citet{Currie.2014}, $^{(12)}$\citet{Bowler.2014}, $^{(13)}$\citet{Luhman.2009}, $^{(14)}$\citet{Wahhaj.2011}, $^{(15)}$\citet{Bejar.2008}, $^{(16)}$\citet{Ireland.2011}, $^{(17)}$\citet{Allers.2005}}
\begin{tabular}{c c c c c c}
\hline\hline
Name$^{ref}$ & SpT & d &  Age & Comp. mass & Sep. \\ 
			&	& (pc)  &   (Myr) &   (M$_\mathrm{Jup}$)   &  (AU) \\ [0.5ex]
\hline
Planetary-mass companions & & & & & \\ [0.5ex]
\hline
2M~0103(AB)~b$^{(1)}$ & M5,5 + M6 & $47.2 \pm 3.1$ & $30$ & $13.0 \pm 1.0$ & $84.0$\\
GU~Psc~b$^{(2)}$ & M3 & $48.0 \pm 5.0$ & $100 \pm 30$ & $11.0 \pm 2.0$ & $2000 \pm 200$\\
J0219-3925~B$^{(3)}$ & M6 & $39.4 \pm 2.6$ & $30-40$ & $12-15$ & $156 \pm 10$ \\
FW~Tau~b$^{(4)}$ & M4 & $145.0 \pm 15.0$ & $1.8^{+0.2}_{-1.0}$ & $10.0 \pm 4.0$ & $330 \pm 30$\\
DH~Tau~b$^{(5)}$ & M0,5 & $\sim$ 140 & $1.0 \pm 0.9$ & $11.0_{-0.003}^{+0.01}$ & $330$\\
2M~044144~b$^{(6)}$ & M8,5	& $140.0$ & $1.0$ &	$7.5 \pm 2.5$ & $15.0 \pm 0.6$\\
CHXR~73~b$^{(7)}$ & M3,25 & $\sim$ 1.6 & $2$ & $12.0_{-5.0}^{+8.0}$ & $200$\\
2M1207~b$^{(8)}$ & M8 & $52.4 \pm 1.1$ &	 $8 \pm 3$ & $4.0_{-1.0}^{+6.0}$ & $46.0 \pm 5.0$\\
VHS~1256-1257~B$^{(9)}$ & M7.5 & $12.7 \pm 1.0$ & $150-300$ & $11.2^{+9.7}_{-15}$ & $102 \pm 9$\\
Ross~458(AB)~c$^{(10)}$ & M2 & $114.0 \pm 2.0$ & $475 \pm 325$ & $8.5 \pm 2.5$ & $1168.0$\\
ROXs~42B~b$^{(11)}$ & M0 & $135.0$ & $2 \pm 1$ & $10.0 \pm 4.0$ & $140.0 \pm 10.0$\\
\hline
Brown dwarfs & & & & & \\ [0.5ex]
\hline
1RXS~J034231.8+121622~B$^{(12)}$ & M4V & $23.9 \pm 1.1$ & $60-300$  & $35 \pm 8$ & $19.8 \pm 0.9$\\
FU~Tau~b $^{(13)}$ & M7,25 &	$140.0$ & $1.0$ & $15.0$ & $800$\\ 
CD-35~2722~b$^{(14)}$ & M1V & $21.3 \pm 1.4$ & $100 \pm 50$ & $31.0 \pm 8.0$ & $67.0 \pm 4.0$\\
GJ~3629~B$^{(12)}$ & M3V & $22 \pm 3$ & $30_{-13}^{+30}$ & $46\pm 6$ & $6.5\pm 0.5$\\
2MASS~J15594729+4403595~B$^{(12)}$ & M2 & $33 \pm 4$ & $120$ & $43\pm 9$ & $187\pm 23$\\
UScoCTIO~108~b$^{(15)}$ & M7 & $145.0 \pm 2.0$ & $11 \pm 2$ & $16.0_{-2.0}^{+3.0}$ & $670.0$\\		
GSC~6214-210~b$^{(16)}$ & M1 & $145.0 \pm 14.0$ & $11 \pm 2$ & $17.0 \pm 3.0$ & $320.0$\\
Oph~11~b$^{(17)}$ & M9 & $145.0 \pm 20.0$ & $11 \pm 2$ & $21.0 \pm 3.0$ & $243 \pm 55.0$\\
1RXS~J235133.3+312720~B$^{(12)}$ & M2 & $50 \pm 10$ & $120$ & $32.0 \pm 6.0$ & $120 \pm 20$\\
\end{tabular}
\label{Mcompanions}
\end{center}
\end{table*}

Both populations of Planetary Mass Companions (hereafter PMCs, $<14M_\mathrm{Jup}$; close-in RV and transit PMCs on one hand, and wide-orbit DI PMCs on the other) challenge our view on planetary formation. Core accretion \citep[CA,][]{Pollack.1996, Alibert.2004} followed by planet migration is the preferred model to explain both the formation of Solar System giant planets and Hot Jupiters found around solar-type stars. Nevertheless, CA does not easily explain the massive gas giants imaged around low-mass stars, especially at large separations, even in the case of pebble accretion \citep{Lambrechts.2012}. Indeed, the mass of the planets are of the same order of magnitude (e.g. 2M0103b, \citealt{Delorme.2013}) or even greater (2M1207b, \citealt{Chauvin.2004}) than the total mass of the protoplanetary disc ($\approx$~10\% of the star's mass). On the other hand, Gravitational Instability \citep[GI,][]{Boss.2011,Cameron.1978} provides an interesting alternative to CA to explain the formation of wide-orbit, massive planets within the disk of a low-mass star. Other initial conditions for the formation of PMCs, such as the position of the ice line and the stellar metallicity, impact their formation mechanisms. For instance, there are indications that planets with masses between 10~$M_\mathrm{Earth}$ and 4~$M_\mathrm{Jup}$ orbiting metal-poor stars are found at larger separations \citep{Adibekyan.2013}. 
\\
Studying low-mass stars provides an opportunity to test the formation of PMCs at the low end of the stellar mass function. They also allow the study of a wider range of companion to star mass ratios. Different mechanisms might explain the formation of the planets, as a function of their mass ratio. In this paper, we will investigate whether there are different populations of planets orbiting low and high-mass stars. 
\\
From an observational point of view, \citet{Bonfils.2011} found that the rate of close-in giant planets ($m \sin(i)=100-1000~M_\mathrm{Earth}$) around M-type stars is low ($f<0.01$ for $P=1-10~d$, $f=0.02^{+0.03}_{-0.01}$ for $P=10-100~\mathrm{d}$), while close-in super-Earths are much more numerous ($f=0.35^{+0.45}_{-0.11}$ for $P=10-100d$). Concerning higher-mass stars, \citet{Rameau.2013} find that between 10.8\% and 24.8\% of A to F type stars host at least one planet defined by the parameter intervals [1,13]~$M_\mathrm{Jup}$ and [1,1000]~AU. Recently, \citet{Bowler.2014} showed that few M dwarfs host giant planets closer than for <1000~AU and argue that there is no evidence of a relation between the wide-orbit giant planet (>10~AU) frequency and the stellar mass (for A, FGK and M stars). However, other studies combining different techniques of detection (RV and DI, micro lensing and RV) have shown that GPs are less frequent around M dwarfs \citep[respectively]{Montet.2014,Clanton.2014}.
\\
In this paper, we present the results of the M-dwArf Statistical Survey for direct Imaging of massiVe Exoplanets (MASSIVE). Our sample is composed of 58 young and nearby M dwarfs observed in $L^\prime$-band for which we have follow-up data for all our candidate companions.
\\
\\
We present the MASSIVE survey in Section~\ref{MASSIVE_survey}. Section~\ref{Frequency} details our Bayesian approach to derive the frequency of planetary companions orbiting low-mass stars. In Section~\ref{Bayesian_Analysis}, we explore the influence of the stellar host mass on the planet occurrence by comparing the planetary companion frequency around low-mass stars in the MASSIVE survey and around a similar VLT/NaCo survey that targeted higher mass, A-F stars \citep{Rameau.2013}. We present our conclusions in Section~\ref{Conclusion}. In all the following Sections, we consider three types of companions: PMCs with masses $<14M_\mathrm{Jup}$, brown dwarfs with masses between $14$ and $80M_\mathrm{Jup}$, and SCs that bring together PMCs and brown dwarfs ($<80M_\mathrm{Jup}$).

\section{The MASSIVE survey}
\label{MASSIVE_survey}
\subsection{Observations}

\subsubsection{Sample selection}
58 M-dwarfs have been observed. This sample was built as follows:  
\\
- \textbf{Age}: We selected our targets from the known members of young stellar associations \citep[e.g.][]{Zuckerman.2004}. We selected 10 stars from TW~Hydrae \citep[5-15~Myr; we use 8~Myr, hereafter TWA,][]{Kastner.1997,Weinberger.2013,Webb.1999}, 22 stars from the $\beta$~Pictoris moving group \citep[20-26~Myr; we use 21~Myr, hereafter BPMG,][]{Binks.2014,Malo.2014, Mamajek.2014}, 8 stars from Tucana-Horologium \citep[20-40~Myr; we use 30~Myr, hereafter THA,][]{Zuckerman.2000,Kraus.2014_tucana}, 3 stars from Columba \citep[20-40~Myr; we use 30~Myr, hereafter COL,][]{Torres.2008}, 1 star from Argus \citep[30-50~Myr; we use 40~Myr,][]{Makarov.2000}, and 12 stars from ABDoradus \citep[110-130~Myr; we use 120~Myr, hereafter ABDor][]{Luhman.2005,Barenfeld.2013}. Two other targets are young, but do not belong a stellar association.
\\
The membership of our targets is based on publications listed in Table~\ref{targets}. The number of known young, M-type dwarfs has increased substantially thanks to recent surveys (such as 2MASS, ROSAT, Galex, and WISE) and to efforts to identify and characterize new, low-mass members of young moving groups \citep[e.g.,][]{Torres.2008, Schlieder.2012, Malo.2013, Gagne.2015}.
\\
- \textbf{Distance}:
The M dwarfs in our sample are nearby ($d<62$~pc) so that we  can probe small star-companion projected separations, typically down to $5-10$~AU.
\\
- \textbf{Brightness}:
The sample stars are brighter than $K_S=12$ to allow for good AO correction and for the detection of low-mass planets.  
\\
\\
These criteria are summarized for each target in Table~\ref{targets}. Note that the closest and the youngest stars were observed in higher priority.

\begin{table}
\centering
\caption{Observing periods. V stands for visitor mode, S for service mode.}
\begin{tabular}{c c c c}
\hline\hline
Obs. period & run & mode\\ [0.5ex]
\hline
Dec. 2009 & 084.C-0739 & S \\
Aug. 2012 & 089.C-0665(B) & V \\
Nov. 2012 and Feb. 2013 & 090.C-0698(A) & V \\
Nov. 2012 & 090.C-0728(A) & S \\ 
Aug. 2013 & 291.C-5031(A) & S \\ [1ex]
\end{tabular}
\label{obs}
\end{table}

\subsubsection{The data}
Data were acquired from December 2009 to February 2013, over five runs (see Table~\ref{obs}) with VLT/NaCo \citep{Rousset.2003}. The observing conditions are described in Table~\ref{conditions}. NaCo includes an adaptive optics system (NAOS) and a near-infrared camera \citep[CONICA,][]{Lenzen.2003}. It was mounted at the Nasmyth focus of the 8.2~m telescope UT4 at the time of the observations. CONICA is equipped with a CCD infrared camera, and an Aladdin 3 detector (1024x1024 pixels). We used the available infrared wavefront sensor (IRWFS/NAOS) to close the adaptive optics (AO) loop on faint red stars such as most MASSIVE targets.
\\
We used the L27 camera (27.1 mas/pixel, and a field of view of 28''x28'') and the cube mode of NaCo, recording data-cubes of single frames with exposure times as low as 0.175~s. We observed in $L^{\prime}$ band ($\lambda = 3.8$~$\mu$m, $\Delta \lambda = 0.62$~$\mu$m) to improve our sensitivity  to the coolest and therefore lowest-mass companions.  We used the pupil tracking mode with a high parallactic angle in order to apply Angular Differential Imaging (ADI, \citealt{Lafreniere.2007}, \citealt{Marois.2006}), a method that allows speckle subtraction and deeper close-in sensitivity. Otherwise, we used regular field tracking.

\onecolumn
\begin{landscape}
\begin{longtable}{|c|c|c|c|c|c|c|c|c|c|c|}
\caption{Targets of MASSIVE with their characteristics. }
\\
\hline
ID star & Period & RA & DEC & age & assoc. & d & SpTyp & K & Total Exposure Time & Refs. \\ 
  &  &  &  & (Myr) &  & (pc) & & & (s) & \\ \hline
\endfirsthead
 \caption{continued.}\\
 \hline
 \endhead
 \hline
 \endfoot
2MASS J00243202--2522528 & 089.C-0665(B) & 00:24:32 & -25:22:53 & 120 & ABDor & 41 & $M2$ & 9.0 & 2560 & 8 \\
2MASS J00251465--6130483& 089.C-0665(B) & 00:25:14.7 & -61:30:48.3 & 30 & THA & 45.8 & $M0$ & 7.8 & 4800 & 1,12 \\
2MASS J00255097--0957398 & 291.C-5031(A) & 00:25:51 & -09:57:40 & 120 & ABDor & 31.8 & $M4$ & 9.0 & 2240 & 8 \\
2MASS J00452814--5137339 & 089.C-0665(B) & 00:45:28.1 & -51:37:33.9 & 30 & THA & 40.4 & $M1$ & 7.6 & 3840 & 12 \\
2MASS J01033563--5515561 & 090.C-0698(A) & 01:03:35.6 & -55:15:56.1 & 30 & THA & 47 & $M7$ & 9.2 & 1280 & 2,3\\
2MASS J01071194--1935359 \tablefootmark{*}& 090.C-0728(A) & 01:07:11.9 & -19:35:36 & 21 & BPMG & 43 & $M1$ & 7.3 & 3820 & 5 \\
2MASS J01231125--6921379 & 090.C-0698(A) & 01:23:11.3 & -69:21:37.9 & 30 & THA & 42.2 & $M8.5$ & 11.3 & 1440 & 2,4 \\
2MASS J01365516--0647379 & 090.C-0698(A) & 01:36:55.2 & -06:47:37.9 & 21 & BPMG & 24 & $M4$ & 8.9 & 2880 & 5,12 \\
2MASS J01521830--5950168 & 089.C-0665(B) & 01:52:18.3 & -59:50:16 & 30 & THA & 40 & $M2.5$ & 8.1 & 3840 & 7 \\
2MASS J02224418--6022476 & 090.C-0698(A) & 02:22:44.2 & -60:22:47.6 & 30 & THA & 32 & $M4$ & 8.1 & 1280 & 7,8 \\
2MASS J02365171--5203036 & 084.C-0739 & 02:36:51.5 & -52:03:04.4 & 30 & COL & 39 & $M2$ & 7.5 & 2880 & 1,14 \\
2MASS J03350208+2342356 & 090.C-0698(A) & 03:35:02.1 & +23:42:35.6 & 21 & BPMG & 42 & $M8.5$ & 11.3 & 320 & 4,7 \\
2MASS J03363144--2619578 & 090.C-0698(A) & 03:36:31.4 & -26:19:57.8 & 30 & THA & 44 & $M6$ & 9.8 & 1040 & 2,6 \\
2MASS J03472333--0158195 & 090.C-0698(A) & 03:47:23.3 & -01:58:20 & 120 & ABDor & 16.1 & $M3$ & 6.9 & 1080 & 1,12 \\
2MASS J04141730--0906544 & 090.C-0698(A) & 04:14:17 & -09:06:54 & 120 & ABDor & 24 & $M4$ & 8.8 & 320 & 7,8,11 \\
2MASS J04373746--0229282 & 084.C-0739 & 04:37:37.5 & -02:29:28.2 & 21 & BPMG & 29 & $M0$ & 6.4 & 3840 & 1,5 \\
2MASS J04433761+0002051 & 084.C-0739 & 04:43:37.6 & +00:02:05.2 & 21 & BPMG & 25.7 & $M9$ & 11.2 & 2640 & 2,4,13 \\ 
2MASS J04464970--6034109 & 090.C-0698(A) & 04:46:49.7 & -60:34:10.9 & 40 & Argus & 37 & $M1.5$ & 7.7 & 4320 & 7 \\ 
2MASS J04522441--1649219 & 090.C-0698(A) & 04:52:24.4 & -16:49:22 & 120 & ABDor & 16 & $M3$ & 6.9 & 2520 & 1,7\\ 
2MASS J04533054--5551318 & 089.C-0665(B) & 04:53:30.5 & -55:51:32 & 120 & ABDor & 11 & $M3$ & 6.9 & 2400 & 1,5 \\ 
2MASS J04593483+0147007 & 084.C-0739 & 04:59:34.8 & +01:47:00.7 & 21 & BPMG & 25.9 & $M0$ & 6.3 & 480 & 1,12 \\ 
2MASS J05004714--5715255 & 084.C-0739 & 05:00:47.1 & -57:15:25.5 & 21 & BPMG & 26.8 & $M0$ & 6.2 & 1920  & 1,12 \\ 
2MASS J05015881+0958587 & 090.C-0698(A) & 05:01:58.8 & +09:58:58 & 21 & BPMG & 25 & $M3$ & 6.4 & 840 & 5 \\ 
2MASS J05064991--2135091 & 084.C-0739 & 05:06:49.6 & -21:35:06.0 & 21 & BPMG & 19.2 & $M2$ & 6.1 & 960 & 1,5 \\ 
2MASS J05195412--0723359& 090.C-0698(A) & 05:19:54.1 & -07:23:35.9 & 30 & COL & 43 & $M4$ & 10.2 & 640 & 2,5 \\ 
2MASS J05320450--0305291 \tablefootmark{*}& 090.C-0698(A) & 05:32:04.5 & -03:05:29 & 21 & BPMG & 42 & $M4$ & 7.0 & 2850 & 1,5\\ 
2MASS J06085283--2753583 & 090.C-0698(A) & 06:08:52.8 & -27:53:58.3 & 21 & BPMG & 31 &  $M8.5$ & 12.4 & 1960 & 4,5,10 \\ 
2MASS J06255610--6003273 & 090.C-0698(A) & 06:25:55.9 & -60:03:27.6 & 120 & ABDor & 23 & $M3$ & 7.2 & 3840  & 1\\ 
2MASS J07285117--3015527 & 090.C-0698(A) & 07:28:51.2 & -30:15:52.7 & 120 & ABDor & 15.7& $M5$ & 8.1 & 2560 & 2 \\   
2MASS J07285137--3014490 \tablefootmark{*} & 090.C-0698(A) & 07:28:51.5 & -30:14:47 & 120 & ABDor & 15.7 & $M1$ & 5.7 & 1920 & 1,5 \\ 
2MASS J08173943--8243298 & 090.C-0698(A) & 08:17:39.4 & -82:43:29.8 & 21 & BPMG & 27 & $M3.5$ & 6.8 & 480 & 7 \\  
2MASS J10284580--2830374 & 090.C-0698(A) & 10:28:45.8 & -28:30:37.4 & 8 & TWA & 50 & $M5$ & 10.0 & 2880 & 2,9 \\ 
2MASS J10285555+0050275 & 084.C-0739 & 10:28:55.6 & +00:50:27.6 & 120 & ABDor & 7.1 & $M2.5$ & 5.3 & 1920 & 1,12 \\  
2MASS J11020983--3430355 & 090.C-0698(A) & 11:02:09.8 & -34:30:35.5 & 8 & TWA & 56.4 & $M8.5$ & 11.9 & 1980 & 1,2\\ 
2MASS J11210549--3845163 & 090.C-0698(A) & 11:21:05.6 & -38:45:16 & 8 & TWA & 65 & $M2$ & 8.1 & 3840 & 5,17 \\ 
2MASS J11393382--3040002 & 090.C-0698(A) & 11:39:33.8 & -30:40:00.3 & 8 & TWA & 42 & $M5$ & 9.1 & 1920 & 2,9 \\ 
2MASS J11395113--3159214 & 084.C-0739 & 11:39:51.1 & -31:59:21.5 & 8 & TWA & 28.5 & $M8$ & 11.5 & 6480 & 1,2 \\  
2MASS J12072738--3247002 \tablefootmark{1} & 090.C-0698(A) & 12:07:27.4 & -32:47:00 & 8 & TWA & 54 & $M1$ & 7.8 & 3600 & 7,17 \\ 
2MASS J12073346--3932539& 084.C-0739 & 12:07:33.5 & -39:32:53.9 & 8 & TWA & 52 & $M8$ & 12.0 & 8460 & 1,2 \\ 
2MASS J12153072--3948426 & 084.C-0739 & 12:15:30.7 & -39:48:42.6 & 8 & TWA & 51 & $M1$ & 7.3 & 960 & 1,7,18 \\ 
2MASS J12265135--3316124 & 090.C-0698(A) & 12:26:51.3 & -33:16:12.4 & 8 & TWA & 61.8 & $M4$ & 9.8 & 1440 & 5 \\ 
2MASS J13215631--1052098 & 090.C-0698(A) & 13:21:56.3 & -10:52:09.8 & 8 & TWA & 40 & $M4.5$ & 8.6 & 2400 & 16 \\ 
2MASS J14112131--2119503 \tablefootmark{*}& 090.C-0698(A) & 14:11:21.3 & -21:19:50.3 & [0-1000] & & 28 & $M7.5$ & 11.3 & 1740 & 4,15 \\  
2MASS J15385757--5742273 & 090.C-0698(A) & 15:38:56.9 & -57:42:18 & 21 & BPMG & 40 & $M4$ & 9.4 & 4000 & 1,5 \\
2MASS J16334161--0933116 & 089.C-0665(B) & 16:33:41.6 & -09:33:12 & 120 & ABDor & 30 & $M0$ & 7.6 & 2560  & 1,5 \\ 
FS2003 0979 & 089.C-0665(B) & 18:35:20.8 & -31:23:24.2 & 100 &  & 18 & $M5$ & 8.0 & 990 & 14 \\ 
2MASS J18465255--6210366 & 089.C-0665(B) & 18:46:52.6 & -62:10:36.6 & 21 & BPMG & 54 & $M1$ & 7.9 & 3520 & 1,7 \\ 
2MASS J19560294--3207186 & 089.C-0665(B) & 19:56:02.9 & -32:07:18.7 & 21 & BPMG & 55 & $M4$ & 8.1 & 2560 & 7  \\  
2MASS J19560438--3207376 & 084.C-0739 & 19:56:04.4 & -32:07:37.7 & 21 & BPMG & 55 & $M0$ & 7.9 & 3480 & 7 \\  
2MASS J20333759--2556521 & 089.C-0665(B) & 20:33:37.6 & -25:56:52.2 & 21 & BPMG & 48.3 & $M4.5$ & 8.9 & 3650 & 5,12 \\ 
2MASS J20450949--3120266 & 084.C-0739 & 20:45:09.5 & -31:20:27.2 & 21 & BPMG & 10 & $M1$ & 4.5 & 1040 & 1,12 \\ 
2MASS J20450949--3120266 \tablefootmark{2} & 084.C-0739 & 20:45:09.5 & -31:20:27.2 & 21 & BPMG & 10 & $M1$ & 4.5 & 880 & 1,12 \\ 
2MASS J21100535--1919573 & 089.C-0665(B) & 21:10:05.4 & -19:19:57.4 & 21 & BPMG & 32 & $M2$ & 7.2 & 1920 & 5,12 \\ 
2MASS J21443012--6058389 & 089.C-0665(B) & 21:44:33.1 & -60:58:38.9 & 30 & THA & 44 & $M0$ & 7.9 & 3520 & 1,12 \\ 
2MASS J22445794--3315015 & 084.C-0739 & 22:44:58.0 & -33:15:01.7 & 21 & BPMG & 23.3 & $M1$ & 6.9 & 960 &1,12 \\ 
2MASS J22450004--3315258 & 084.C-0739 & 22:45:00.0 & -33:15:25.8 & 21 & BPMG & 23.3 & $M5$ & 7.8 & 960 & 1,12 \\  
2MASS J23301341--2023271 \tablefootmark{3}& 090.C-0698(A) & 23:30:13.4 & -20:23:27.1 & 30 & COL & 16.2 & $M3$ & 6.3 & 1640 & 5,7 \\ 
2MASS J23323085--1215513 & 084.C-0739 & 23:32:30.9 & -12:15:51.4 & 21 & BPMG & 28 & $M0$ & 6.6 & 960 & 1,7 \\ 
2MASS J23381743--4131037 & 090.C-0698(A) & 23:38:17.4 & -41:31:03.7 & 120 & ABDor & 19 & $M2$ & 7.5 & 2560 & 8 
\label{targets}
\end{longtable}
\tablefoot{
\\
\tablefoottext{1}{This target is a spectroscopic binary \citep{Shkolnik.2011}.}
\\
\tablefoottext{2}{This target has been observed twice, in pupil tracking mode of VLT/NaCo.}
\\
\tablefoottext{3}{This target is a spectroscopic binary \citep{Riedel.2014}.}
\\
\tablefoottext{*}{Stars excluded from our statistical analyses, cf. Section~{Frequency}}
\\
{\bf References.} (1) \citet{Schleider.thesis}; (2) \citet{Gagne.2015}; (3) \citet{Gagne.2014}; (4) \citet{Gagne.2014bis}; (5) \citet{Malo.2013}; (6) \citet{Rodriguez.2013}; (7) \citet{MaloIII.2014}; (8) \citet{Schleider.thesis}; (9) \citet{Schneider.2012}; (10) \citet{Rice.2010}; (11) \citet{Shkolnik.2012}; (12) \citet{Malo.2014}; (13) \citet{Schlieder.2012}; (14) \citet{Chauvin.2010}; (15) \citet{Gagne.2015_VII}; (16) \citet{Riaz.2006}; (17) \citet{Weinberger.2013}; (18) \citet{Ducourant.2014}.
}
\end{landscape}

\twocolumn

\subsection{Data reduction}
\label{data_analysis}
We use our Interactive Data Language (IDL) pipeline developed for AO reduction at the Institut de Plan\'etologie et d'Astrophysique de Grenoble (IPAG) to reduce the VLT/NaCo data. The reduction steps are described in \citet{Lagrange.2010} and \citet{Delorme.2012a}. They include bad pixel removal, flat-fielding, and frame recentering. Two different methods are used to detect faint companions in the data,  dependending on the separation from the star.
\\
At separations closer typically than 0.5'', noise is dominated by quasi-static speckles. To mitigate this and reveal faint, close-in companions, we employ a modified ADI algorithm called Smart ADI \citep[SADI,][]{Marois.2006}. The reference point spread function used for PSF subtraction in the SADI framework were derived from a restricted number of images (10) where each had rotated by at least one Full width at half maximum (FWHM) at a separation of 0.4''. Candidate companions in this region are identified by visual inspection of the residual images.
\\
At separations typically larger than 0.5'', where the noise in $L^\prime$-band is background limited, we apply a median-filtering technique. For each pixel in the region of interest, we subtract the median in a $20 \times 20$~pixel box centered on that pixel. This step is performed on each frame before stacking the images. Candidate companions are identified in this region using the automatic point source detection routine described in \citet{Delorme.2012a}. When the field rotation is too small to apply SADI, which is the case for 18 of our targets, we only apply median-filtering.
\\
By combining these complementary approaches to candidate detection, we maximize our sensitivity at all separations. The boundary of the inner/outer zone that are used for each target are given in Table~\ref{contrasts} and detailed in Section~\ref{detection_boundaries}.

\subsection{Detection results}
\subsubsection{Companion detections}
We detail hereafter the status of two companions already reported in previous papers and that are part of MASSIVE: 2MASS~J01033563--5515561(AB)b and 2MASS~J12073346--3932539b. In addition, we present the peculiar case of 2MASS~J08540240--3051366, an M-type star that we excluded from MASSIVE.
\\
\\
{\bf 2MASS~J01033563--5515561(AB)b}
\\
2MASS~J01033563--5515561(AB), hereafter 2MASS0103(AB), is a 30~Myr M-dwarf binary located at 47~pc. \citet{Delorme.2013} discovered an exoplanet orbiting this target, with an estimated mass of $12-14M_{Jup}$ using BT-Settl models \citep{Allard.2012}. We do not present any new result with respect to this target.
\\
\\
{\bf 2MASS~J12073346--3932539b}
\\
2MASS~J12073346--3932539 (hereafter 2M1207) is an 8~Myr brown dwarf located at 52.4~pc. \citet{Chauvin.2005.2M1207} reported the discovery of an exoplanet orbiting this target.
\\
2M1207b ($4 \pm 1~M_{Jup}$) is at a separation of $772 \pm 4$~mas (PA: $125.4 \pm 0.3^{\circ}$), $768 \pm 5$~mas (PA: $125.4 \pm 0.3^{\circ}$), and $776 \pm 8$~mas (PA: $125.5 \pm 0.3^{\circ}$) from its host star, respectively, for the VLT/NaCo data from 27/04/2004, 05/02/2005, and 31/03/2005. In the 2009 data, we find a separation of $756 \pm 12$~mas, with a position angle of $124.6^{\circ}$. Our new measurement thus overlaps with the previously reported results and no additional contraints on the orbit is added.
\\
\\
{\bf 2MASS~J08540240--3051366B}
\label{J0854_par}
\\
We conducted an in-depth analysis (Appendix~C) of 2MASS~J08540240--3051366 (hereafter 2MASS~J0854) because we identified a potentially interesting bound companion around it. However, a deeper analysis, using a new radial velocity measurement, revealed that the system is likely older than suspected (several Gyr instead of 10~Myr) and that the companion is a low-mass star. Thus, we do not include this target in our analysis.

\subsubsection{Background stars}
\begin{figure*}[h]
\begin{center}
\begin{tabular}{c}
	\includegraphics[width = 150mm]{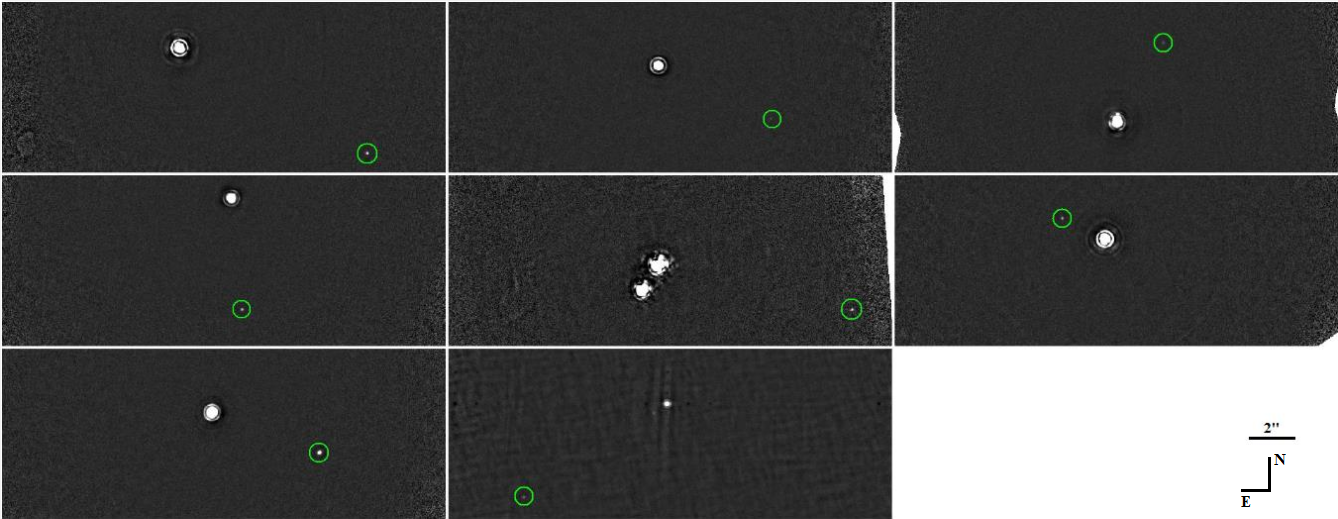} \\
	\includegraphics[width = 50mm]{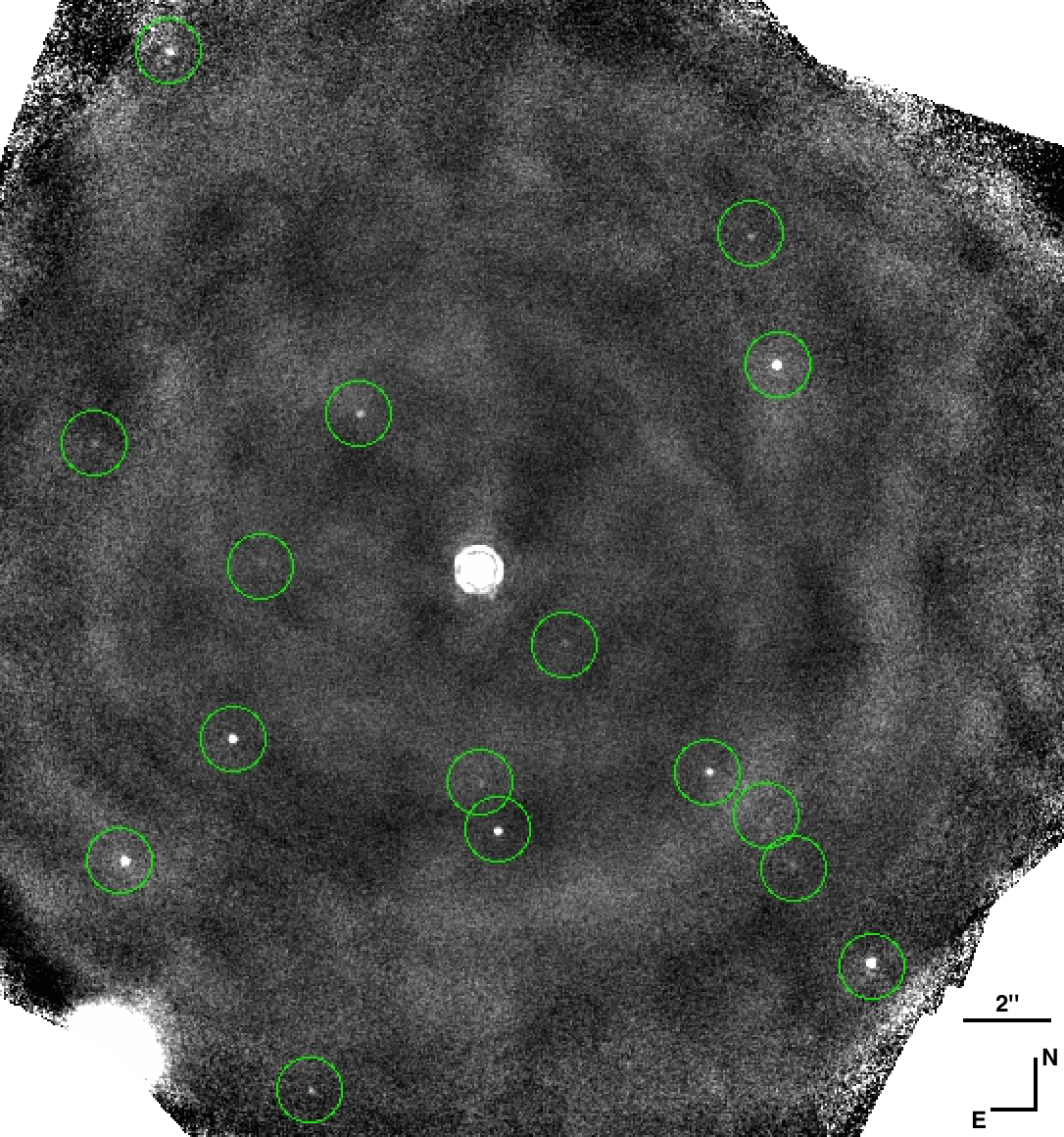}
\end{tabular}
\caption{Background stars identified in the MASSIVE survey, highlighted with green circles. The primary M dwarfs are, from left to right and top to bottom: 2MASS~J06255610-6003273, 2MASS~J11210549-3845163, 2MASS~J18465255-6210366, 2MASS~J01365516-0647379, 2MASS~J05015881+0958587, 2MASS~J04522441-1649219, 2MASS~J07285117-3015527, 2MASS~J12073346-3932539, and 2MASS~J15385757--5742273 (the image of this target is at the very bottom). Images are displayed using identical linear scalings. East is left and North is up.} 
\label{contaminants}
\end{center}
\end{figure*}

Nine MASSIVE targets were identified with candidate companions, in addition to 2MASS0103(AB) and 2M1207. Two methods were used to discriminate background objects from bona-fide companions: 
\begin{enumerate}
\item We used data obtained at different epochs, to test whether or not the candidate and the star shared the same proper motion. If they do, this would mean that they are gravitationally bound. 
\item We used data obtained at different wavelengths when it was possible ($J$,$H$,$K_s$,$L^\prime$-bands), to compare the colours of the candidate with that expected for a background contaminant or a young planet companion. Indeed, young low gravity companions have much redder colours than main sequence contaminants: for instance, at 20~Myr, the BT-Settl models predict that a 5~$M_{Jup}$ planet would have $H-L'\approx2.5$, while one of the reddest contaminants such as a 5~Gyr 0.3~M$_{\odot}$ star, would have $H-L'\approx0.6$.   
\end{enumerate}
We find that all of them are background stars. The targets with background stars are: 2MASS~J06255610--6003273, 2MASS~J11210549--3845163, 2MASS~J18465255--6210366, 2MASS~J01365516--0647379, 2MASS~J05015881+0958587, 2MASS~J04522441--1649219, 2MASS~J07285117--3015527, 2MASS~J12073346--3932539, 2MASS~J15385757--5742273 (see Figure~\ref{contaminants}).

\subsubsection{Visual binary stars}
\begin{table*}[h]
\begin{center}
\caption{Close binary systems (<1\textquotedbl) identified in our survey.}
\begin{tabular}{c c c c c}
\hline
 & ID & approx. separation & $\approx \Delta L^\prime$ & Ref. \\ 
  & & (\textquotedbl) & (mag)  & \\ [0.5ex]
\hline
&2MASS J01033563--5515561 & $0.25$ & 0.22 & \\ 
&2MASS J01071194--1935359 & $0.45$ & 0.09 & \\ 
&2MASS J01231125--6921379 & $<0.07$  \tablefootmark{a} & &\\ 
&2MASS J03363144--2619578 & $<0.12$ \tablefootmark{a} & &\\ 
&2MASS J04373746--0229282 & $0.26$ & 1.26 & \citet{Kasper.2007} \\ 
&2MASS J05015881+0958587 & $1.37$ & 1.39 & \citet{Elliott.2015} \\ 
&2MASS J05195412--0723359 & $0.72$ & 0.51 & \\ 
&2MASS J05320450--0305291 & $0.18$ & 0.63 & \\ 
&2MASS J07285137--3014490 & $<0.12$ \tablefootmark{a}  & & \citet{Daemgen.2007} \\ 
&2MASS J08173943--8243298 & $0.62$ & 1.00 & \\ 
&2MASS J13215631--1052098 & $<0.11$ \tablefootmark{a} & \\ 
&2MASS J14112131--2119503 & $<0.15$ \tablefootmark{a} & \\ 
&2MASS J19560294--3207186 & $0.20$ & 1.04 & Elliott et al. (in prep.) \\ [1ex] 
\end{tabular}
\label{etoiles_binaires}
\end{center}
\tablefoot{
\tablefoottext{a}{stars' point spread function not resolved}
}
\end{table*} 
We identified 13 close binary (<1\textquotedbl) stars, which are reported in Table~\ref{etoiles_binaires}. Some of them are very well resolved while others are tight or suspected binaries (some data show elongated intensity distributions).These binaries will be studied within a larger sample of young binaries in another paper \citep{Bonavita.2016}.

\subsection{Detection limits in mass and semi-major axis for MASSIVE}
In this section, we present the performances of MASSIVE in a [mass, semi-major axis] parameter space.

\subsubsection{2D contrast maps}
Since we apply a different reduction technique at close and at large separations to detect SCs (see section~\ref{data_analysis}), we also derive 2D detection contrast maps in each case  \citep[see ][]{Delorme.2012a}. Note that since the focus of this article is a statistical analysis, we choose to present the results from the simple and robust ADI technique such as SADI, rather than the slightly more sensitive LOCI \citep{Marois.2006} or PCA \citep{Soummer.2012} analyses that use subjective parameters (such as the size of the optimization zones in LOCI or the numbers of components kept in PCA) that can be optimized for individual analyses but not easily adapted for such a systematic analysis. Note that the benefit of LOCI or PCA in sensitivity are moderated in $L^\prime$-band because the speckle-dominated area is smaller. We therefore caution that our detection limits on individual stars are probably conservative at short separations (see Table~\ref{contrasts}) by typically a few tenths of magnitudes with respect to the best performance that could be achieved with a carefully tuned LOCI or PCA analysis.

\subsubsection{2D mass detection limit maps}
\label{detection_boundaries}
\begin{figure}
\begin{center}
\includegraphics[width = 80mm]{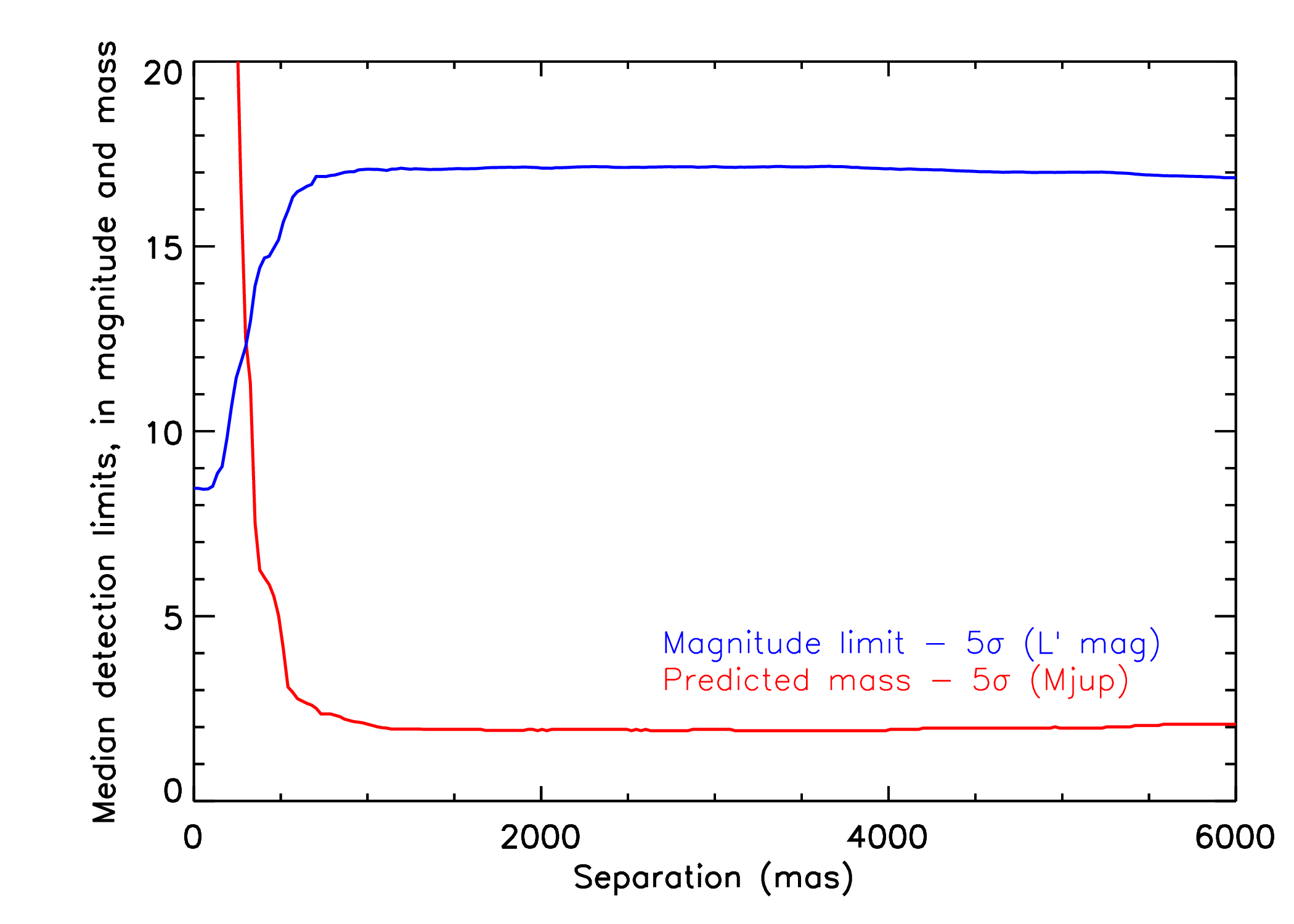}
\caption{Median detection performances for MASSIVE in magnitude ($L^\prime$-band) and mass ($M_\mathrm{Jup}$).}
\label{detlimmass}
\end{center}
\end{figure}

We use \citet{Baraffe.2003} evolution models that assume a hot start together with BT-Settl model atmosphere \citep{Allard.2012} to estimate the masses of planets detectable at 5$\sigma$ given the computed contrast limits. We emphasize that the systematic errors of the models are very difficult to quantify, and can be significant, see e.g. \citet{Dupuy.2014}. However, hot-start models have been commonly used for a long time and provide the best reference for the comparison between various imaging surveys. 
\\
Since the inner and outer regions of the 40 stars where SADI was applied are dominated by different sources of noise, we generate composite maps to cover the full range of possible semi-major axes. We determine the location of the boundary between the inner and outer regions on a star by star basis by comparing the detection limits reached using each image processing technique (the brighter the target, the wider the boundary, see Table~\ref{contrasts}).
\\
Figure~\ref{detlimmass} gives the median detection performances of MASSIVE in terms of contrasts in $L^\prime$-band and mass as a function of semi-major axis. The median detection performances for MASSIVE in apparent magnitudes of companions, is $L^\prime \approx$17~mag at 1'' (at 5$\sigma$) and further away. The contrasts reachable at $L^\prime$ are presented in Table \ref{contrasts}. The MASSIVE survey is the deepest DI survey yet conducted on M dwarfs.

\subsubsection{Detection probabilities}
\label{detec_probabilities}
\begin{figure*}
   \begin{minipage}[c]{.50\linewidth}
      \includegraphics[width = 95mm]{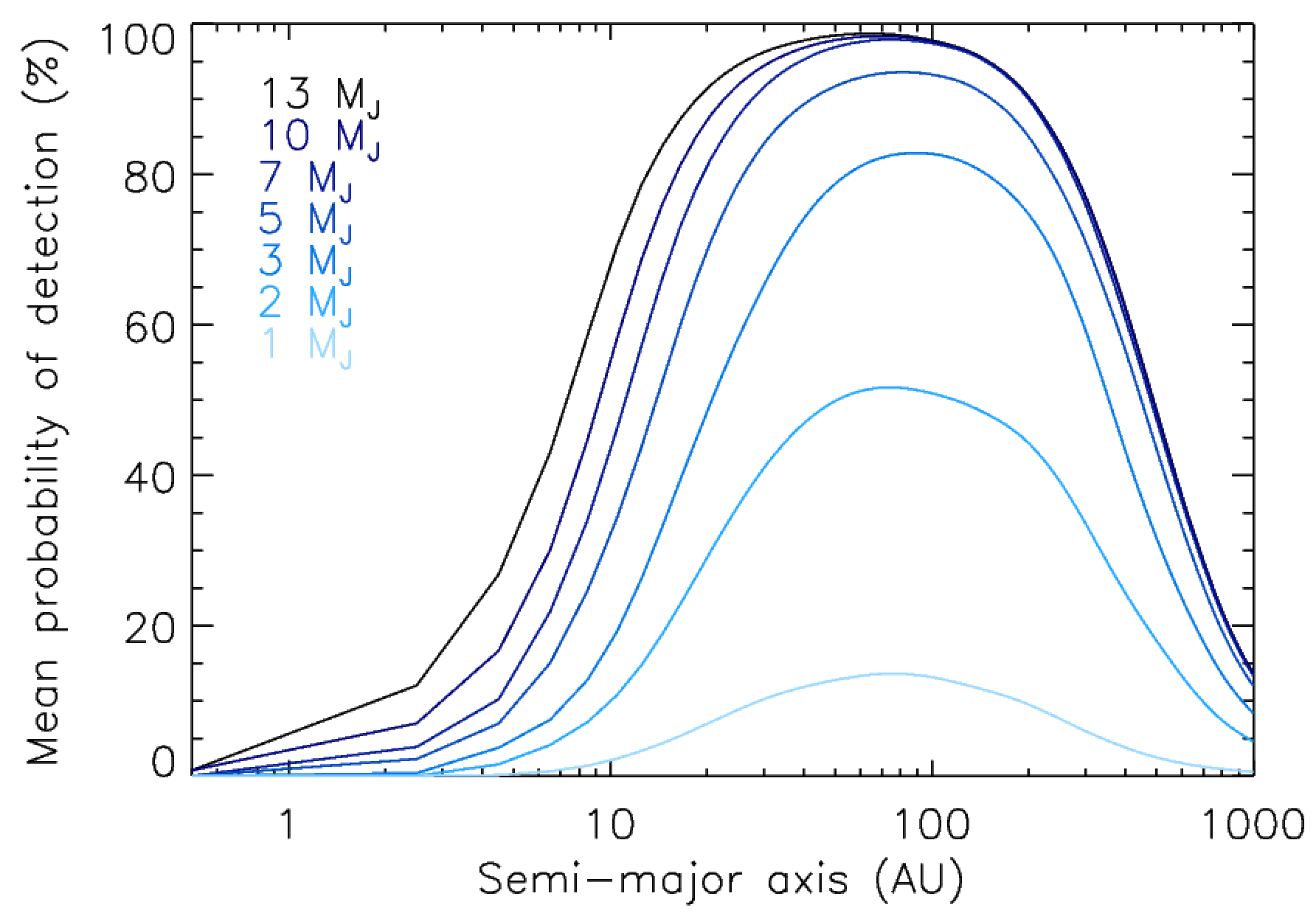}
   \end{minipage} \hfill
   \begin{minipage}[c]{.50\linewidth}
      \includegraphics[width = 95mm]{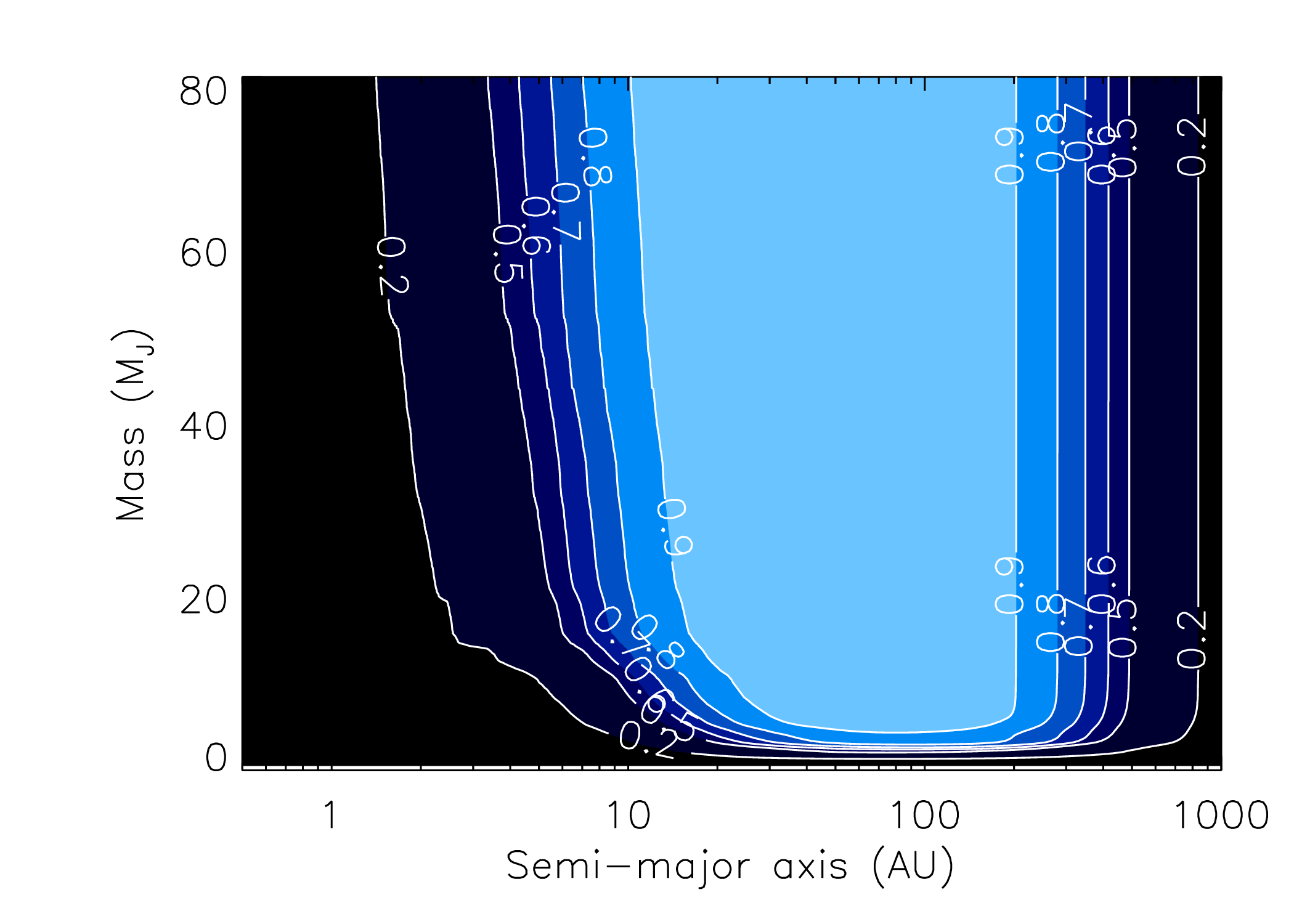}
   \end{minipage}
   \caption{MASSIVE survey mean detection probability curves (left) and contour plot of detection probabilities (right) for several masses, as a function of semi-major axis.}
   \label{meanproba}
\end{figure*}

Using the 2D mass detection limit maps that we built in Section~\ref{detection_boundaries}, we estimate the probability to detect companions orbiting our stars, as a function of given mass and semi-major axis.
\\
We use an optimized version \citep{Rameau.2013} of the original MESS code \citep{Bonavita.2011} which takes into account projection effects that can hide a companion behind its host star on a fraction of its orbits. Hence the detection limits represent the sensitivity to planets according to their semi-major axis (hereafter, SMA) and not to their projected separations. From a technic point of view, MESS is a Monte Carlo simulation code that builds synthetic planet populations described in terms of frequency, orbital elements and physical properties. It uses a semi-empirical approach to generate the planet populations using a regularly spaced grid. We generate $100000$ planets for each [mass,SMA] pair of the grid. The semi-major axis range is $0.5$ to $1000~\mathrm{AU}$ with steps of $2~\mathrm{AU}$ and the masses are in the range [0.5,80]~$M_\mathrm{Jup}$ with steps of 0.16~M$_{Jup}$. The MESS code computes the SMA for each synthetic planet for a given mass and period. It also assigns eccentricities, inclinations, longitudes of periastron, longitudes of ascending node and times of periastron passage for each orbit, assuming uniform distributions. The eccentricity values marginally affect the detection probabilities, while the inclination strongly affects the detection probabilities.
\\
We apply the MESS code to all of our targets, taking into account the host star age, distance, and mass (using BT-Settl models), as well as the masses and semi-major axes of the simulated planets, to derive mean planet detection probabilities for all MASSIVE targets (see Fig.~\ref{meanproba}).
\\
We found that MASSIVE has a good sensitivity (mean detection probability $> 50$\%) in the range 8 to 400~AU and 2 to 80~M$_\mathrm{Jup}$ (see Fig.~\ref{meanproba}) from the generation of $100000$ planets for each point of the [mass,SMA] grid. We consider only this best parameter range in most of the following analysis.

\section{Frequency of planetary companions orbiting low-mass stars of the MASSIVE survey}
\label{Frequency}
We detail the statistical formalism used for our Bayesian analysis in Appendix~A.
\\
For this analysis (Sections~\ref{Frequency} and \ref{Bayesian_Analysis}), we exclude 4 targets out of the 58 targets of the MASSIVE sample, either because the estimated age is too uncertain (i.e. not members of young moving groups, like it is the case for 2MASS~J14112131--2119503), or because of the close binarity of some targets that makes the total mass too big to consider it as a low-mass system (this is the case for 2MASS~J01071194--1935359, 2MASS~J05320450--0305291 and 2MASS~J07285137--3014490, see Table~\ref{mass_MASSIVE}). However, we kept in our survey components of wide-separation binaries for which our AO observations allowed to look for SCs around each component individually. This led to a homogeneous sample of 54 targets.

\begin{table*}[h]
\begin{center}
\caption{Estimated masses of our MASSIVE targets, using their absolute magnitudes and the BT-Settl models \citep{Allard.2012}.}
\begin{tabular}{c c}
\hline
ID & Mass \\ 
 & (M$_{Jup}$) \\ [0.5ex]
\hline
2MASS J00243202--2522528 & 450 \\
2MASS J00251465--6130483 & 700 \\
2MASS J00255097--0957398 & 400 \\
2MASS J00452814--5137339 & 700 \\
2MASS J01033563--5515561 & 450 \\
2MASS J01231125--6921379 & 65 \\
2MASS J01365516--0647379 & 110 \\
2MASS J01521830--5950168 & 500 \\
2MASS J02224418--6022476 & 400 \\
2MASS J02365171--5203036 & 400 \\
2MASS J03350208+2342356 & 50 \\
2MASS J03363144--2619578 & 230 \\
2MASS J03472333--0158195 & 450 \\
2MASS J04141730--0906544 & 310 \\
2MASS J04373746--0229282 & 820 \\
2MASS J04433761+0002051 & 20 \\ 
2MASS J04464970--6034109 & 620 \\ 
2MASS J04522441--1649219 & 500 \\ 
2MASS J04533054--5551318 & 350 \\ 
2MASS J04593483+0147007 & 750 \\ 
2MASS J05004714--5715255 & 800 \\ 
2MASS J05064991--2135091 & 570 \\ 
2MASS J05195412--0723359 & 150 \\ 
2MASS J06085283--2753583 & 12\tablefootmark{*} \\ 
2MASS J06255610--6003273 & 570 \\ 
2MASS J07285117--3015527 & 250 \\ 
2MASS J08173943--8243298 & 650 \\ 
2MASS J10284580--2830374 & 90 \\ 
2MASS J10285555+0050275 & 450 \\  
2MASS J11020983--3430355 & 250 \\
2MASS J11210549--3845163 & 570 \\ 
2MASS J11393382--3040002 & 150 \\ 
2MASS J11395113--3159214 & 15 \\ 
2MASS J12072738--3247002 & 500 \\
2MASS J12073346--3932539 &  20 \\ 
2MASS J12153072--3948426 & 700 \\ 
2MASS J12265135--3316124 & 350 \\ 
2MASS J13215631--1052098 & 182 \\
2MASS J15385757--5742273 & 175 \\
2MASS J16334161--0933116 & 600 \\ 
FS2003 0979 & 300 \\ 
2MASS J18465255--6210366 & 750 \\ 
2MASS J19560294--3207186 & 820  \\ 
2MASS J19560438--3207376 & 750 \\  
2MASS J20333759--2556521 & 350 \\
2MASS J20450949--3120266 & 620 \\ 
2MASS J20450949--3120266 & 620 \\
2MASS J21100535--1919573 & 600 \\
2MASS J21443012--6058389 & 620 \\ 
2MASS J22445794--3315015 & 450 \\
2MASS J22450004--3315258 & 250 \\ 
2MASS J23301341--2023271 & 500 \\ 
2MASS J23323085--1215513 & 700 \\ 
2MASS J23381743--4131037 & 450 
\end{tabular}
\label{mass_MASSIVE}
\tablefoot{
\tablefoottext{*}{The association of this star \citep[BPMG,][]{Riedel.2014,Malo.2013} were debated by \citet{Gagne.2014bis} who show that this target could be a member to the Columba association, but other radial velocity measurements are still needed to confirmed this. The mass of this object is quite low here, and is more certainly around 15-24~M$_{Jup}$, as reported by \citet{Riedel.2014} and \citet{Gagne.2014bis}. We however kept this low mass in order to have a consistent determination of all the stellar masses. }
}
\end{center}
\end{table*}

\subsection{Optimal parameter range to be considered in our formalism}
\label{Optimal parameter range}
The choice of parameter range on which we carry out our statistical analysis has a direct impact on the results. For instance, if we choose a broad range, such as [1-13]~$M_\mathrm{Jup}$ and [1-1000]~AU used by \citet{Rameau.2013}, we include probability integration ranges like [1-8]~AU or [400-1000]~AU where our probability of detecting any planet is close to zero. This would lead to larger errors because of the lack of actual information injected into the probability calculation. Also, since we carried out a Bayesian analysis, including such ranges with no observational constraints would artificially bias the results toward a posterior that is close to the supposed prior. We therefore choose to focus on ranges of mass and SMAs for which the MASSIVE observational constraints are stronger (cf. section~\ref{detec_probabilities}), i.e. [2,80]~$M_\mathrm{Jup}$ or [1,80]~$M_\mathrm{Jup}$ at [8,400]~AU.

\subsection{Results of our Bayesian analysis}
\subsubsection{Working hypotheses}
We choose to address several distinct statistical questions (hereafter, we use $Q$ as the mass ratio between a host star and its planet):
\begin{itemize}
\item What is the probability to have a massive companion 
(2~M$_{Jup}<$M$<$80~$M_\mathrm{Jup}$) regardless of its $Q$ value ? ({\bf case 1})
\item What is the probability to have a massive companion 
(2~M$_{Jup}<$M$<$80~$M_\mathrm{Jup}$) at an intermediate mass ratio $0.01 \le Q \le 0.05$ ? ({\bf case 2})
\item What is the probability to have a massive companion 
(1~M$_{Jup}<$M$<$80~$M_\mathrm{Jup}$) at a low mass ratio $ Q \le 0.01$ ? ({\bf case 3})
\end{itemize}
We choose an upper limit of 0.05 for the intermediate mass ratio (case 2), because it excludes most low-mass star companions. Such companions might make the system appear as a binary, which could cause a selection bias because known binaries were excluded in the survey of \citet{Rameau.2013}. Choosing such an upper limit in mass ratio keeps only the very high contrast companions in the selection range, and therefore effectively removes this possible bias. 
\\
We choose 1~$M_\mathrm{Jup}$ as a lower mass limit for case~3 since our previous lower mass limit (2~$M_\mathrm{Jup}$, cases~1 and 2) is not relevant for such low $Q$ because in practice it excludes all the lowest mass stars (<0.2~M$_{\odot}$) of our sample. Indeed, by choosing 1~$M_\mathrm{Jup}$ instead of 2~$M_\mathrm{Jup}$, we increase the effective number of LMSs (number of stars in sample times average companion detection probability in the chosen semi-major axis and mass range) by 15\% (see Tab.~\ref{effective_number}). So, decreasing the lower-mass limit allows to better constrain case~3, which is not true for cases~1 and 2. Note that our statistical formalism correctly addresses the fact that the sensitivity of MASSIVE concerning some stars for case~2 and 3 is very weak or even zero, with the hypothesis that the planetary frequency in the considered mass range is homogeneous. 
\begin{table*}
\centering
\caption{Effective number of stars.}
\begin{tabular}{c c c c c}
\hline\hline
Survey & Case 1 & Case 2 & Case 3 \\ [0.5ex]
\hline \hline
MASSIVE & 45.9 & 40.0 & 23.2 \\
A-F survey & 27.6 & 28.3 & 23.9\\
\hline
\end{tabular}
\label{effective_number}
\end{table*}
\\
\\
More precisely, our three cases correspond to 3 working hypotheses:
\\
{\bf Case 1.} We consider all detections, regardless of their $Q$: we want to know the frequency of stars hosting substellar companions (PMCs and brown dwarfs). We will therefore consider all known detections, namely the systems 2MASS0103(AB) and 2M1207.
\\
{\bf Case 2.} We want to know the frequency of stars hosting SCs having 
a moderate mass ratio $Q$ (that is $0.01<Q<0.05$) only. With the 
limitations detailed below (see \ref{comp_freq}), this intermediate $Q$ range would 
tentatively  target objects formed via GI. For this second case, there 
is one detection in our sample (2MASS0103(AB)b). The median value for the MASSIVE targets is 0.45~M$_{\odot}$, corresponding to 22.5~M$_{Jup}$ and 4.5~M$_{Jup}$ companion upper mass limit for $Q=0.05$ and $Q=0.01$ respectively.  
\\
{\bf Case 3.} We consider SCs with a very low mass ratio Q, $Q \le 0.01$. With the limitations detailed below (see \ref{comp_freq}), this very low $Q$ range tentatively targets objects that more likely formed via CA. There are no detections of such low-$Q$ companions in the MASSIVE survey.

\subsubsection{Companion frequencies for low-mass stars in MASSIVE.}
\label{comp_freq}
\begin{figure*}
\includegraphics[width = \linewidth] {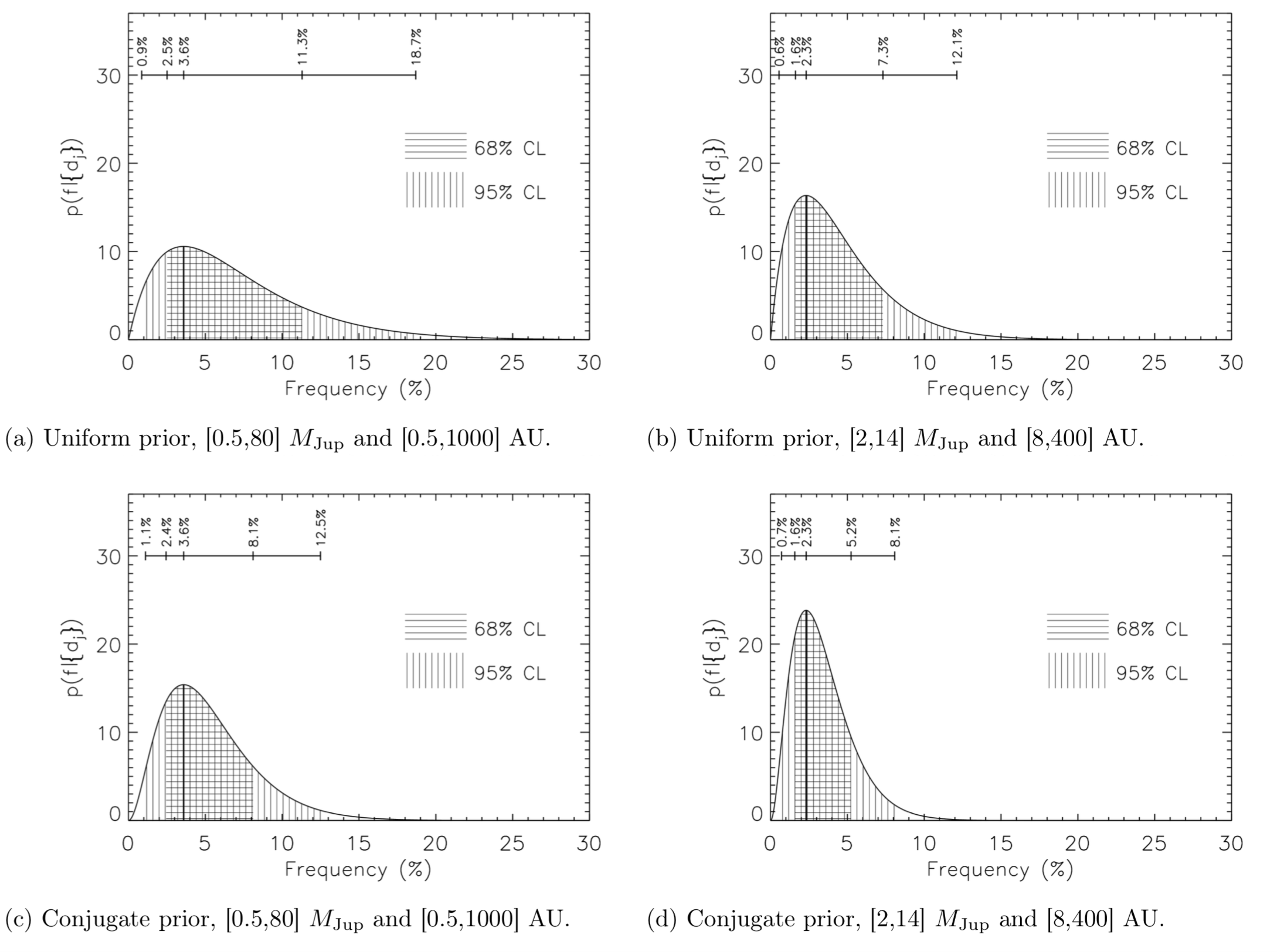}
  \caption{Fractions of low-mass stars in MASSIVE hosting SCs (left) and PMCs (right) for different [mass, SMA] ranges, and priors. We take into account the only detection of a PMC around 2MASS0103(AB).}
  \label{proba1}
\end{figure*}

\begin{figure*}
\includegraphics[width = \linewidth] {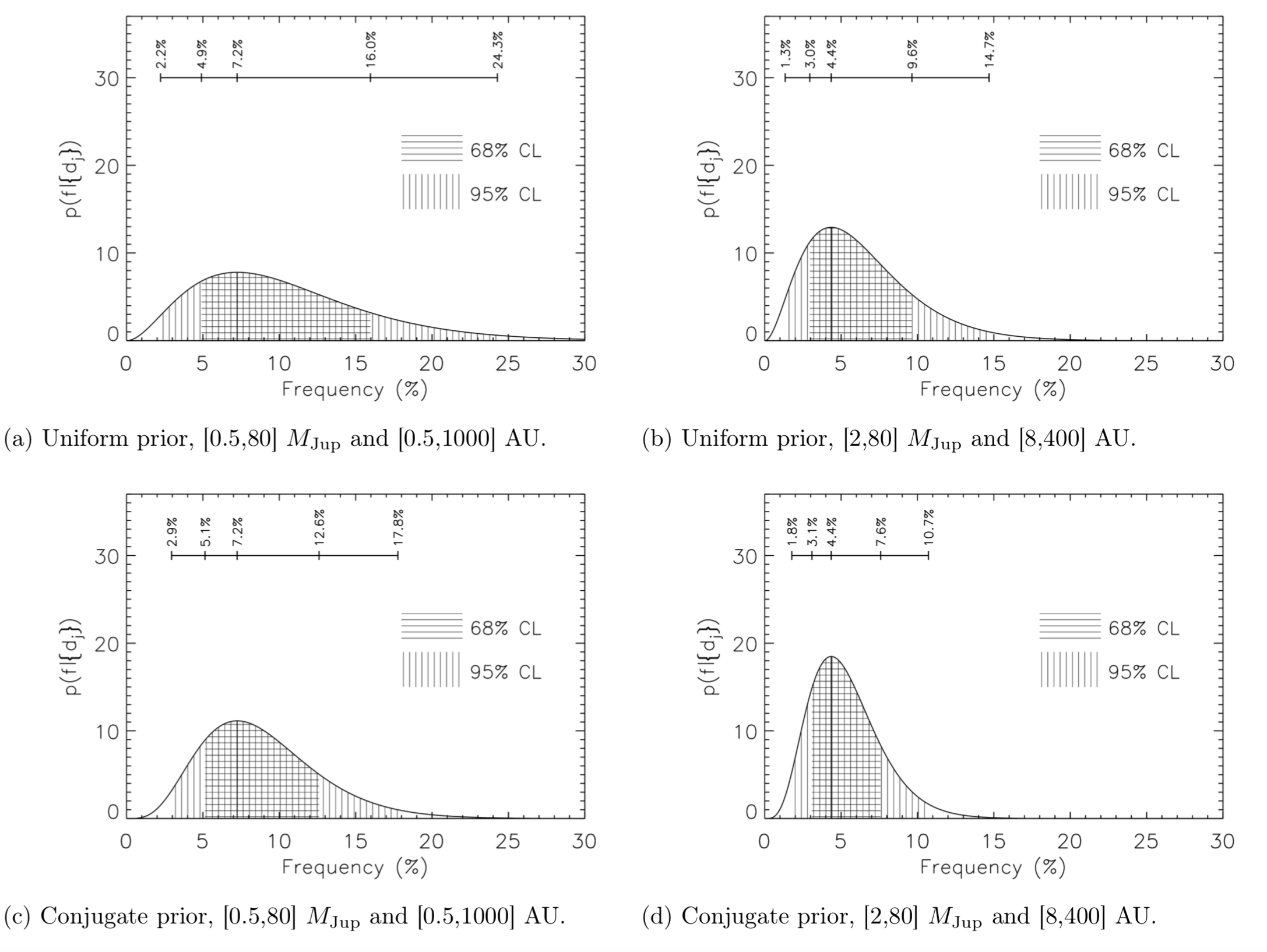}
  \caption{Fractions of low-mass stars in MASSIVE hosting SCs for different [mass, SMA] ranges, and priors. We take into account the detection of a PMC around 2MASS0103(AB) and 2M1207.}
  \label{proba2}
\end{figure*}

We use Equations $(.4)$ and $(.5)$ to determine the minimum and maximum value of the substellar companion frequency given a confidence level, respectively $f_{min}$ and $f_{max}$, using Bayes rule and a binomial likelihood distribution (see Appendix~A). In the case where no companion is detected in a given parameter range, we use Equation $(.9)$ to determine $f_{max}$. We use both natural conjugate priors and uniform priors. Using a uniform prior is a strong hypothesis in terms of planet frequency probability distribution, and a hypothesis that is rather discrepant with the relatively low observed giant planet frequency. We nevertheless use it to compare our results with published ones which generally use it, mostly because uniform prior does not use any observational information, in other words it is fixed regardless of the data. Note also that the flat prior is not an uninformative prior. We also use our observational constraints by estimating a natural conjugate prior (see Appendix~A) in our empirical Bayesian analysis. We consider that this conjugate prior is better adapted to our analysis than the uniform prior because it is much more consistent with observations.
\\
\\
Our statistical results concerning the MASSIVE survey are reported in Table~\ref{MESS}, Fig.~\ref{proba1} and \ref{proba2}. The figures show the frequency distributions derived considering the whole mass and SMA ranges, ie [0.5,80]~M$_\mathrm{Jup}$ and [0.5,1000]~AU. These parameter ranges are not optimal for our analysis (cf. Section~\ref{Optimal parameter range}) but present the advantages to show how sensitive the frequency distributions are with respect to the parameters choice. The results that we report in the next sections are based on the use of optimized parameters ranges (cf. Section~\ref{Optimal parameter range}). We find that $2.3^{+2.9}_{-0.7}$~\% (68\% confidence level CL) of our low-mass stars host at least one GP in the mass range 2-14~$M_\mathrm{Jup}$ and SMA range 8-400~AU. To derive this value, we use the natural conjugate prior and took into account the detection of the 2MASS0103 system. When we consider the same SMA range and a broader mass range, i.e. 2-80~$M_\mathrm{Jup}$, we find that $4.4^{+3.2}_{-1.3}\%$ of our low-mass stars host at least one SC (68\% CL, case 1), using a natural conjugate prior and considering the detection of 2MASS0103(AB)b and 2M1207b. The frequency distributions are wider when considering the uniform prior instead of the conjugate prior (see Table~\ref{MESS}, Fig.~\ref{proba1} and \ref{proba2}): the lower value does not change significantly, which is due to the fact that uniform priors assume more numerous planets, while conjugate priors use the observations that revealed few companions.
\\
Our frequencies are lower for cases 2 and 3 which is expected because cases 2 and 3 companions  are subsamples of case1 companions. We derive that $2.5^{+3.1}_{-0.8}$~\% of the low-mass stars host at least one SC (68\% CL, case 2: 2~M$_{Jup}<$M$<$80~$M_\mathrm{Jup}$ and $0.01 \le Q \le 0.05$), the fraction range is [0\%,2.6\%] for case 3 (1~M$_{Jup}<$M$<$80~$M_\mathrm{Jup}$ and $ Q \le 0.01$).

\subsubsection{Strengths and limitations of our approach}
Our objective is to derive the planet frequency around low-mass stars and compare this value with the frequency of planets orbiting high-mass stars (see Section \ref{Bayesian_Analysis}). In the present analysis, we focus on the mass ratio $Q$ between the planets and their host stars instead of their masses, since the resulting $Q$ can significantly differ if considering low-mass or high-mass stars. For instance, while $Q$ is around 20\% for the 2M1207 system, its value is less than 1\% for the $\beta$~Pictoris system \citep{Lagrange.2010}, more than an order of magnitude smaller, while the companion masses are comparable. This issue is particularly important since planetary formation mechanisms that take place in a protostellar disc (such as CA and GI) directly depend on the disc mass and hence on the stellar mass \citep{Andrews.2013}. Formation scenarios are probably more sensitive to $Q$ than to the mass of the companion. Indeed, CA is not expected to form planets with a mass ratio greater than a few percents \citep[e.g.]{Mordasini.2012}. In contrast, GI has difficulties to form low $Q$ companions and tends to form only companions with higher $Q$, such as brown dwarfs or even low-mass stars \citep[e.g.][]{Dodson-Robinson.2009,Vorobyov.2013}.
\\
\\
However, the mass ratio needs to be used with caution, because it sometimes corresponds to companion mass ranges where survey sensitivities 
are low (e.g for $Q$<0.01, see "case 3" below, around low-mass stars). Another shortfall of 
using the mass ratio to extract observational constraints on companion 
formation mechanisms is that even though the range where each 
formation mechanism takes place is qualitatively clear (CA at low $Q$, GI at moderate 
$Q$ and stellar formation mechanisms for high $Q$), they would likely overlap \citep[e.g.][]{Reggiani.2016}.
Nevertheless, the semi-arbitrary limits in the $Q$ interval 
we retain for "case 2" (0.01<$Q$<0.05) accommodate both our goals of 
probing the range of "intermediate $Q$", and of probing the range of mass where our 
observational constraints are relevant. Moreover, in case 2, where $Q=0.05$ and the host star mass is 2~M$_\odot$ (the typical mass of an A-type star), we should include companions of up to 100~$M_\mathrm{Jup}$ that are cut by our additional limit in absolute mass of 80$M_\mathrm{Jup}$. Thus, the fraction $f$ of the most massive stars hosting SCs is under-estimated for case 2. That limitation cannot be avoided when comparing planet imaging surveys of low-mass stars and more massive hosts whether or not using mass ratio. Moreover, depending on the mass ratio, 
the statistical significance of each observation (that is the 
probability to detect a companion) is taken into account, depending 
on the considered mass ratio. For instance, a young late-M dwarf with a 
mass of 0.1~$M_{\odot}$ would have a high statistical significance in case 
1 because contrast around faint stars is 
usually favorable. On the contrary, this late-M dwarf would have a very low significance for case 3 ($Q$<0.01) 
because only planets lower than 1~$M_\mathrm{Jup}$ would be taken into account, and our observations are only marginally sensitive to these masses. Despite these limitations, using the $Q$ approach is an improvement over the common approach that consists of using only masses.

\begin{table*}
\centering
\caption{Detections of PMCs or brown dwarfs for MASSIVE and the A-F type star surveys, relative to our cases. We selected the median values of the stellar and companion masses to estimate the case each system belongs to.}
\begin{tabular}{c c c c c}
\hline\hline
System & Case 1 & Case 2 & Case 3 & Ref. \\ [0.5ex]
\hline \hline
{\bf MASSIVE} &  &  & & \\
2MASS~J01033563-5515561 & $\checkmark$ & $\checkmark$ & & \citet{Delorme.2013}\\ 
2MASS~J12073346-3932539 & $\checkmark$ & & & \citet{Chauvin.2004}\\ 
\hline
{\bf A-F survey} &  &  & & \\
$\beta$~Pictoris & $\checkmark$ & & $\checkmark$ & \citet{Bonnefoy.2014}\\ 
HR~8799 & $\checkmark$ & & $\checkmark$ & \citet{Pueyo.2014}\\ 
HIP~95261 & $\checkmark$ & $\checkmark$ & & \citet{Lowrance.2000}\\ 
\hline
\end{tabular}
\label{Q}
\end{table*}

\section{Impact of the stellar mass on the planet frequencies.}
\label{Bayesian_Analysis}
\begin{figure}
\begin{minipage}[c]{.50\linewidth}
\includegraphics[width = 90mm]{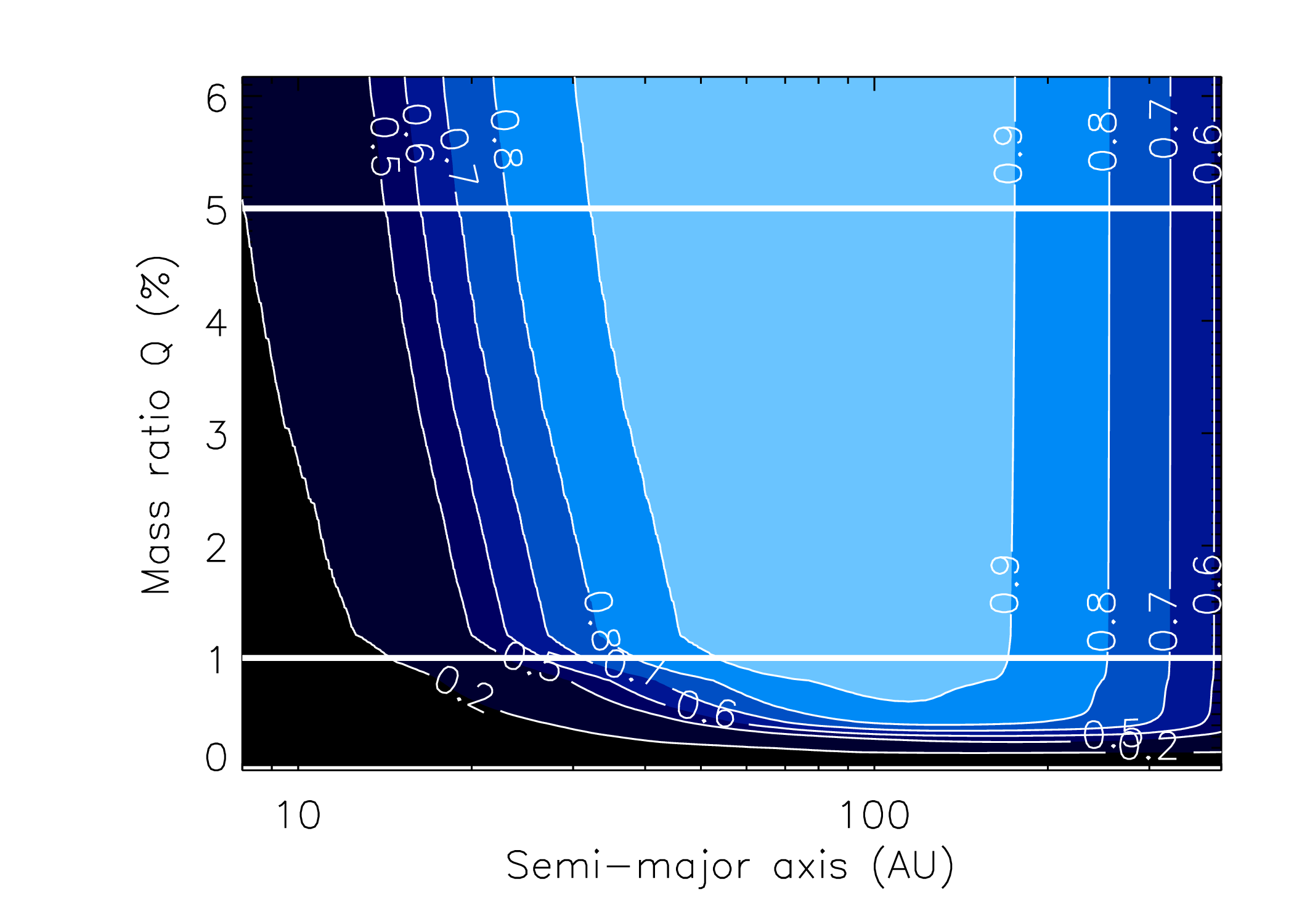}
\end{minipage} \hfill   
\begin{minipage}[c]{.50\linewidth}
\includegraphics[width = 90mm]{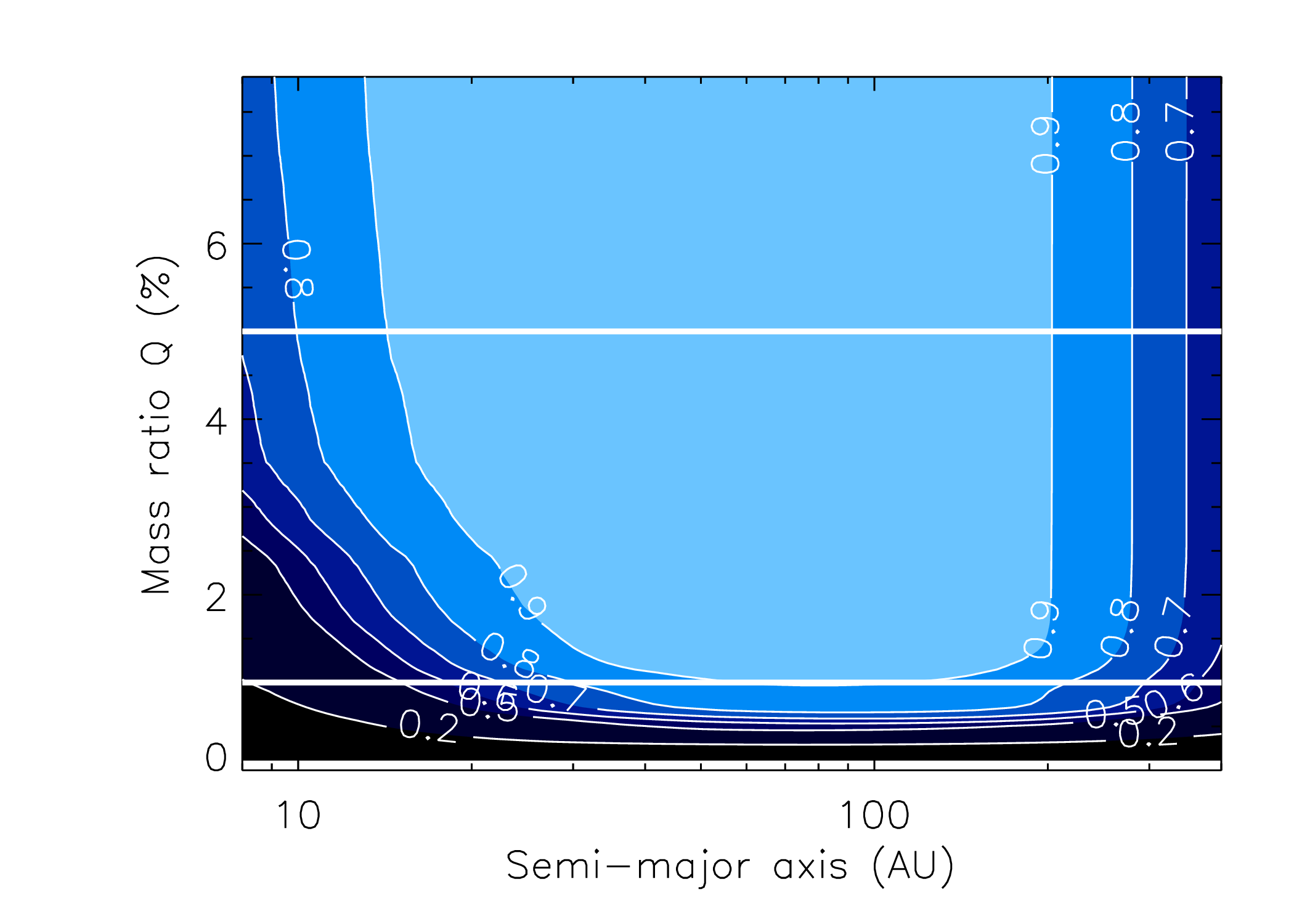}
\end{minipage} \hfill  
\caption{Detection probability contours for the A-F survey (top) and MASSIVE (bottom) for several masses, as a function of semi-major axis. The two white lines correspond to the mass ratio limits for cases 2 and 3 (respectively $0.01<Q<0.05$ and $0.01>Q$).}
\label{Q_compare}
\end{figure}

In this Section, we test the dependence of planet occurrence rate with respect to stellar mass (to identify possible population differences). To do so, we compare the planet frequencies found for the low-mass stars in MASSIVE to the ones corresponding to higher mass stars in \citet{Rameau.2013}, which we derive using the same statistical formalism. \\
The sample that we use for comparison is made of 37 A to F type stars that are young (<100~Myr), nearby (<65~pc), and bright (K<7~mag)\footnote{The star around which a new planet has been imaged \citep[51~Eri~b][]{Macintosh.2015} belongs the A-F stars survey. However, the analysis led by \citet{Rameau.2013} takes into account the non-detectability of this planet (with the 2013 data) in the frequency derivation, so that this value is not affected by this new discovery.}.  
\\
Both surveys (MASSIVE and A-F type stars survey) share roughly the same sample selection criteria (distance, age). The data were obtained, reduced and analyzed with the same tools so that relative biases are limited. The sensitivity of each survey is different which is properly handled by our statistical formalism see Fig.~\ref{Q_compare}). In the following, we refer the MASSIVE targets as low-mass stars (hereafter LMS), and the A-F type stars as high-mass stars (hereafter HMS).

\subsection{LMS and HMS companion frequency comparisons.}
\label{LMS and HMS companion frequency comparisons.}
\begin{figure}
\includegraphics[width = 90mm]{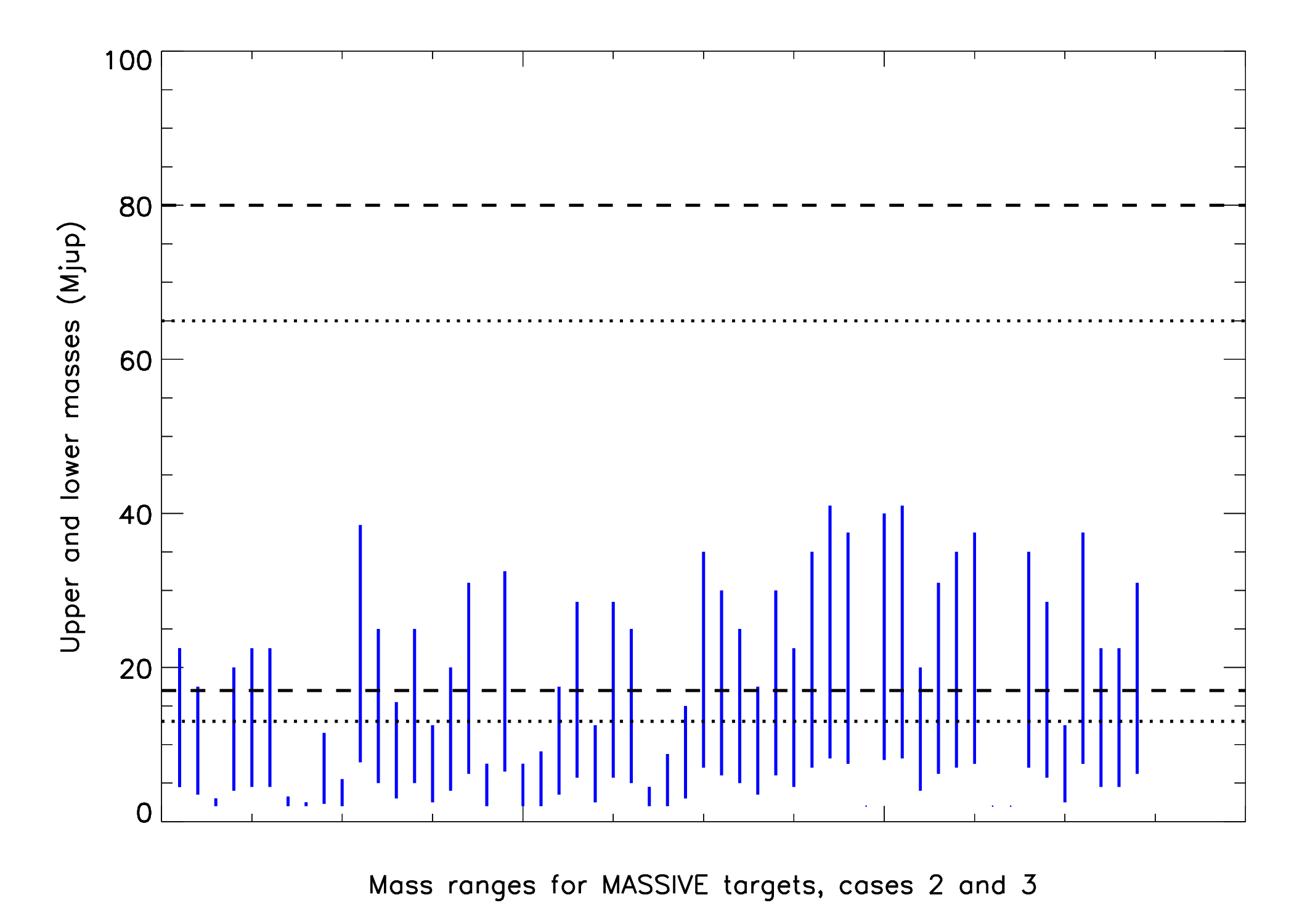}
\caption{Representation of the mass ranges considered for each of our MASSIVE targets (each vertical line represents one target). The upper dashed (respectively dotted) lines represent the upper mass limit for typical A (1.7~M$_{\odot}$) (respectively F, 1.3~M$_{\odot}$) type stars, and the lowest lines are the lower mass limits for these typical A--F type stars. We limit our study to the planet and brown dwarf mass domain, we therefore fix the upper mass limit to 80~M$_{\odot}$. } 
\label{Q_repart}
\end{figure}

As a first step, we use the Kolmogorov-Smirnov (KS) test to establish if the probability distributions of the SC frequency derived from each survey are similar or not. If the KS statistic is high, it hints that the two observed companion frequency distributions originate from distinct populations. Such a difference would be indicative of different companion populations as a function of the stellar mass. Note that we use here the KS test to compare probability distributions that artificially count thousands of stars that KS test considers as "observations". This means that the KS approach is only qualitatively relevant and cannot be used to provide quantitative probabilities. Deriving such quantitative probabilities is the purpose of the next Section. 
\\
{\bf Case 1.} Actual detections corresponding to this case are made around 2MASS0103(AB) and 2M1207 for the LMS survey, and $\beta$~Pictoris, HR8799 and HIP95261 for the HMS survey (cf. Fig. \ref{Q}). We find (see Fig.~\ref{proba3}) that $4.4^{+3.2}_{-1.3}\%$ (at 68\% CL) of low-mass stars host at least one SC, while this fraction is $10.8^{+5.7}_{-2.9}\%$ for HMS \footnote{The results we obtain for the HMS survey give a lower planetary frequency than those obtained by \citet{Rameau.2013}. Using the same data, they found a probability of $16.1^{+8.7}_{-5.3}$\% at 68\% CL and considering PMCs ([1,13]~$M_\mathrm{Jup}$) at [1,1000]~AU, and they considered two detections, while we find a lower frequency ($10.8^{+5.7}_{-2.9}$\% for case 1). However, the [mass,semi-major axis] domain we consider focuses only on the range of parameters where the MASSIVE sensitivity is high. Moreover the error bars on our derived frequencies are smaller since we used a natural conjugate prior in our statistical formalism.}. This difference might be significant, as it is associated to a high KS statistic (0.71).
\\
{\bf Case 2.} We consider the detection of two systems; 2MASS0103(AB) for the LMS survey and HIP95261 for the HMS survey. We find (see Fig.~\ref{proba3}) that the 68\% confidence companion frequency ranges are [1.7\%,5.6\%] for LMS and [2.4\%,7.9\%] for HMS. We find a KS statistic of 0.23. In Figure~\ref{Q_repart}, we summarize the mass ranges considered in case 2, and the overlap between the LMS the HMS surveys: for some stars, there is very little overlap or no overlap at all, depending on their mass. This is the case for the late M dwarfs in MASSIVE, but our statistical formalism properly handles this, hence they are included even if their actual statistical contribution is marginal or null. 
\\
{\bf Case 3.} No companion in the LMS survey fulfills the conditions of case 3, and for the HMS survey we include two systems ($\beta$~Pictoris and HR8799). We find that both distributions are well separated (see Fig.~\ref{proba3}) with a KS statistic of 0.82. We find that $9.1_{-2.7}^{+6.5}\%$ of the HMS host at least one companion (with 68\% CL), while less than $2.6\%$ (68\% CL) of the LMS host one companion.

\subsection{Relation between the SC frequency and the stellar mass}
\label{relation_frequency_host}
Results presented in Section~\ref{LMS and HMS companion frequency comparisons.} seem to qualitatively hint at the existence of two distinct populations of SCs around LMS and HMS. In this Section, we quantitatively investigate the following question: are there two distinct populations of SCs, one for low-mass stars, and one for more massive stars ?
\\
\\
To mitigate possible observational biases that would artificially enhance the inferred difference, we developed a Monte~Carlo code that uses contrapositive logic to provide a conservative absolute probability of whether there are two distinct populations of SCs, depending on the mass of the host star. The idea is to test if the KS statistics that we derive using our observations when comparing the MASSIVE and the A-F type star surveys are caused by the low number of detected planets, or truly distinct populations. Our rationale is the following: we assume that there is one unique underlying population of SCs by merging the two samples of stars to derive a single frequency distribution. Then, with the MC code, we test how likely it is that the random fluctuations in the observation of this single population can yield frequency distributions as different as our observations. The details of our code are described in Appendix~B.
\\
\\ 
We report our results in Table\ref{MESS_bis}. We find a probability of 74.2\% that there are two populations of SCs in case 1, 20.6\% considering case 2, and 74.5\% for case 3. 

\subsection{Discussion}
\begin{figure*}
\includegraphics[width = \linewidth] {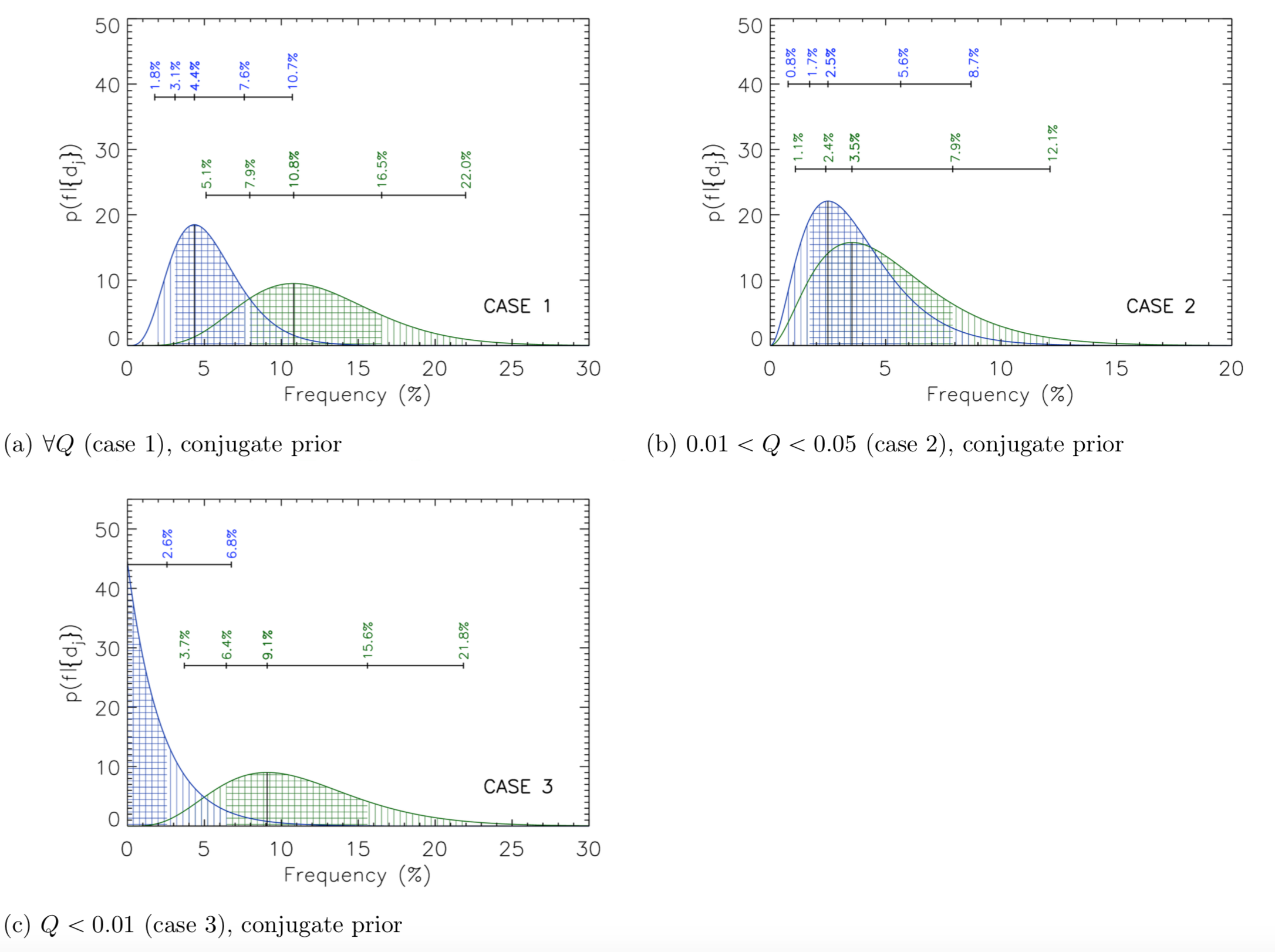}
  \caption{Comparisons of the frequency distributions as function of stellar mass calculated for the three cases described in the text, considering low-mass (blue) and higher mass (green) stars.}
  \label{proba3}
\end{figure*}

We compared the frequencies derived from the two surveys in the 3 different cases of the allowed star-planet mass ratio $Q$ described above.
\\
{\bf Case 1} ( any mass ratio is allowed). We derived SC frequencies of $4.4^{+3.2}_{-1.3}\%$ for the LMS star sample and $10.8^{+5.7}_{-2.9}\%$ for the HMS sample described in \citet{Rameau.2013}. We found that in a single population there is only $\sim$26\% chance to measure frequencies as distinct as we have. This translates to a $\sim$74\% probability that these observations are not randomly drawn from a single population but represent two distinct planet population frequencies.
\\ 
{\bf Case 2} ($0.01<Q<0.05$). We tentatively associated the intermediate mass ratios of Case 2 to planets more likely formed by GI. When comparing the low-mass and high-mass star SC frequencies, we found moderately strong statistical evidence that there are not distinct populations around low-mass and high-mass stars (see Fig.~\ref{proba3}), since we show that there is a $\sim$79\% probability that a unique SC population could reproduce the observed distribution of SCs. However, the corresponding companion masses for low-mass and high-mass stars seem to be quite different; planetary mass companions for LMSs and mostly brown dwarfs for HMSs. Our results for case 2 hints the existence at large separations of an intermediate mass-ratio population of substellar companions that is almost absent in results of surveys probing short separations and whose frequency would not be correlated with stellar mass. However, note that our contrapositive logical approach starts by assuming the existence of such a single population. Therefore, the probability derived for the existence of two populations are conservative by construct (as for cases 1 and 3), while on the opposite the $\sim$79\% of a unique population for case 2 is not. Our results only suggest that the existence of two distinct populations is not necessary to explain the data considered for case 2.
\\
{\bf Case 3} ($Q \le 0.01$). We found a $\sim$75\% probability that two distinct populations of companions exist around LMSs and HMSs. 
\\
\\
We note that given the sensitivity of our survey, we cannot exclude the possibility that less massive giant planets would be as frequent around LMSs as around HMSs. It is known that low-mass planets such as Neptune-like planet masses that form the bulk of gas-giant population around LMSs at short separations \citep[see e.g.][]{Bonfils.2013}, and a substantial population of super-Earths to Jupiter-mass planets are detected at moderate separations ($\sim$~0.5--10~AU) around LMSs using micro-lensing method \citep{Cassan.2012}. Theoretical models of CA also indicate that the formation of giant planets at small separations by CA is difficult around LMSs \citep{Mordasini.2012}. If the same is true at large separations, then the frequency derived in this work are significantly lower than the actual frequency when the Neptune-like planets are also considered.
\\
\\
\\
{\bf Comparison with previous studies}
\\
We used a conjugate Bayesian prior which qualitatively assumes that since we found few giant exoplanets in the range we selected, low frequencies of exoplanets are more probable than high frequencies. This approach is different from that of \citet{Bowler.2015} because 100\% of our targets have been followed-up, providing perfect completeness in our subsequent statistical analysis which is not the case in \citet{Bowler.2015}. In contrast with this study, we found some statistical evidence for a correlation between the SC frequency and the stellar mass at large separations.
\\
It is difficult to compare our results with those from other surveys because most of the time the parameter ranges considered are not exactly the same. We therefore applied our formalism by considering exactly the same domains of mass and semi-major axis as those probed by each of RV and DI \citep{Montet.2014}, micro lensing and RV \citep{Clanton.2014}, and RV surveys \citep{Bonfils.2013}. Table~\ref{comparison_surveys} shows our companion frequencies for these ranges and those derived within these surveys. For our calculations, we considered a 0.5~$M_\mathrm{Sun}$ star and a inclination of 60\degr. Our results are consistent with results described by \citet{Clanton.2014} and \citet{Bonfils.2013}. They reveal a tension at 1.5$\sigma$ with the results of \citet{Montet.2014}, that is of marginal significance.

\begin{table*}
\centering
\tiny
\caption{Comparison of the results from different surveys on M-dwarfs.}
\begin{tabular}{c c c c c}
\hline\hline
Survey & Planet frequency & Range considered & Our results within the same range\\ [0.5ex]
\hline \hline
RV+DI, (1) & 6.5 $\pm$ 3~\% & [1-13]~$M_\mathrm{Jup}$, [0-20]~AU &  [0-3.0]\% \\ 
micro lensing+RV, (2) & 3.8$^{+1.9}_{-2.0}$~\% & [1-13]~$M_\mathrm{Jup}$, $1-10^5$ day periods & [0-2.0]\% \\ 
RV, (3) & <1\% & $m\sin i =[10^3-10^4]~M_\mathrm{Earth}$, $10^3-10^4$ day periods & [0-3.2]\%\\ 
\hline
\end{tabular}
\label{comparison_surveys}
{\bf References.} (1) \citet{Montet.2014}; (2) \citet{Clanton.2014}; (3) \citet{Bonfils.2013}.
\end{table*}

\begin{table*}
\centering
\caption{Frequencies of PMC and SC with $68\%$ confidence level, for different [mass, semi-major axis] ranges, using different priors, for specific detections. The number of detections is represented in the last column: the number $1$ is for the PMC detection around the target 2MASS0103(AB), the number $2$ is for the PMCs detection around 2MASS0103(AB) and 2M1207b.}
\begin{tabular}{c c c c c c}
\hline\hline
Frequency & Interval for 68$\%$ CL & Mass & SMA & prior & number of detections\\ 
                  &                                      & ($M_\mathrm{Jup}$) & (AU) &  & \\ \hline
 & & MASSIVE & & & \\ [0.5ex]
\hline
$3.6\%$ & [2.5,11.3]$\%$ & [0.5,80]& [0.5,1000] & uniform & 1 \\
$3.6\%$ & [2.4,8.1]$\%$ & [0.5,80] & [0.5,1000] & natural conjugate & 1 \\
$2.3\%$ & [1.6,7.3]$\%$ & [2,14] & [8,400] & uniform & 1 \\
$2.3\%$ & [1.6,8.1]$\%$ & [2,14] & [8,400] & natural conjugate & 1 \\
$7.2\%$ & [4.9,16.0]$\%$ & [0.5,80] & [0.5,1000] & uniform & 2\\
$7.2\%$ & [5.1,12.6]$\%$ & [0.5,80] & [0.5,1000] & natural conjugate & 2 \\
$4.4\%$ & [3.0,9.6]$\%$ & [2,14] & [8,400] & uniform & 2\\
$4.4\%$ & [3.1,7.6]$\%$ & [2,14] & [8,400] & natural conjugate & 2 \\
\end{tabular}
\label{MESS}
\end{table*}

\begin{table*}
\centering
\caption{Kolmogorov-Smirnov statistics and associated probability for each case, and the probability of two populations based on a Monte-Carlo simulation.}
\begin{tabular}{l c c}
\hline\hline
Mass ratio Q range & KS statistic & Probability of two populations\\ [0.5ex]
\hline \hline
Case 1, $\forall Q$ & 0.71 & 74.2\% \\ 
Case 2, $0.01 \le Q \le 0.05$ & 0.23 & 20.6\% \\ 
Case 3, $Q<0.01$ & 0.82 & 74.5\% \\ 
\end{tabular}
\label{MESS_bis}
\end{table*}

%

     \begin{figure*}
\includegraphics[width = \linewidth] {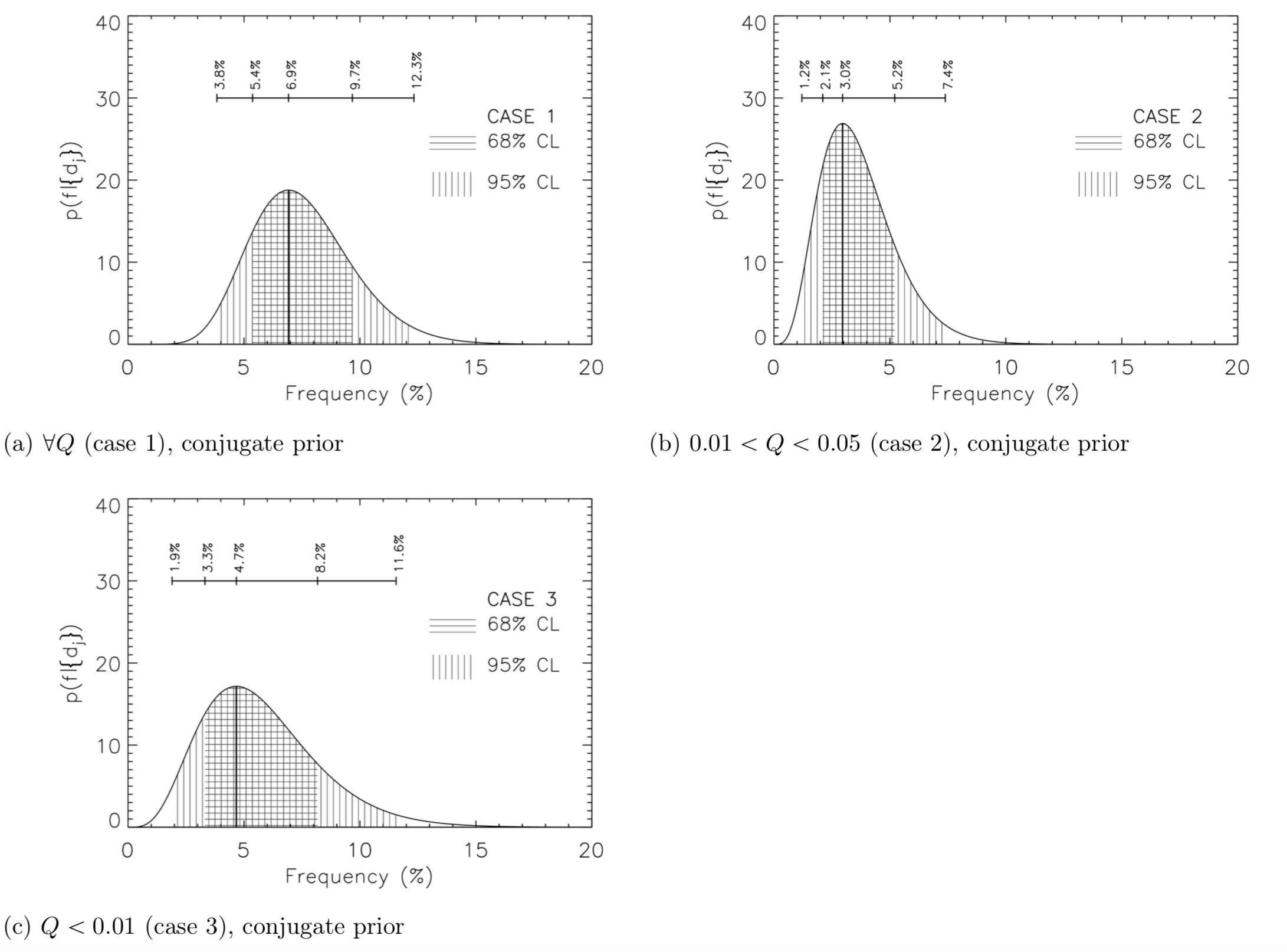}
  \caption{Companion frequency distribution for the three cases considered in the text derived from the merging of the LMS and the HMS samples.}
  \label{proba_2surveys}
\end{figure*}

\begin{table*}
\begin{center}
\caption{Observing conditions.}
\begin{tabular}{|c|c|c|c|c|c|c|c|c|}
\hline
Target & Airmass & Parang. & Tau0 (ms) & Seeing \\
           &               & ($\degr$) & (ms)  &  \\ [0.5ex]
\hline
2MASS J00243202--2522528 & 1.0 & 20.9 & 1.1 & 1.20 \\
2MASS J00251465--6130483 & 1.3 & 27.0 & 1.5 & 0.94 \\
2MASS J00255097--0957398 & 1.0 & 44.2 & 1.9 & 0.99 \\
2MASS J00452814--5137339 & 1.1 & 30.8 & 1.0 & 1.30 \\
2MASS J01033563--5515561 & 1.2 & 8.4 & 2.4 & 0.69 \\
2MASS J01071194--1935359 & 1.0 & 69.9 & 2.2 & 1.00 \\
2MASS J01231125--6921379 & 1.4 & 10.4 & 2.1 & 0.81 \\
2MASS J01365516--0647379 & 1.1 & 30.4 & 1.6 & 1.13 \\
2MASS J01521830--5950168 & 1.2 & 32.5 & 1.3 & 1.01 \\
2MASS J02224418--6022476 & 1.2 & 10.8 & 1.8 & 1.08 \\
2MASS J02365171--5203036 & 1.1 & 25.9 & 2.7 & 1.34 \\
2MASS J03350208+2342356 & 1.5 & 2.0 & 2.8 & 0.60 \\
2MASS J03363144--2619578 & 1.0 & 9.0 & 2.8 & 0.64 \\
2MASS J03472333--0158195 & 1.1 & 10.5 & 2.2 & 0.84 \\
2MASS J04141730--0906544 & 1. & 5.2 & 2.4 & 0.72 \\
2MASS J04373746--0229282 & 1.1 & 30.6 & 4.7 & 1.06 \\
2MASS J04433761+0002051 & 1.1 & 17.5 & 8.9 & 0.83 \\
2MASS J04464970--6034109 & 1.3 & 33.0 & 3.8 & 0.93 \\
2MASS J04522441--1649219 & 1.0 & 56.3 & 2.8 & 0.65 \\
2MASS J04533054--5551318 & 1.2 & 22.1 & 1.4 & 0.98 \\
2MASS J04593483+0147007 & 1.2 & 0.0 & 5.0 & 0.94 \\
2MASS J05004714--5715255 & 1.2 & 14.7 & 3.8 & 0.87 \\
2MASS J05015881+0958587 & 1.2 & 8.9 & 2.3 & 0.76 \\
2MASS J05064991--2135091 & 1.2 & 0.0 & 2.8 & 0.94 \\
2MASS J05195412--0723359 & 1.1 & 5.7 & 5.5 & 1.16 \\
2MASS J05320450--0305291 & 1.1 & 33.2 & 2.1 & 0.84 \\
2MASS J06085283--2753583 & 1.0 & 40.2 & 2.1 & 0.88 \\
2MASS J06255610--6003273 & 1.2 & 16.9 & 7.0 & 0.77 \\
2MASS J07285117--3015527 & 1.0 & 65.8 & 2.6 & 0.70 \\
2MASS J07285137--3014490 & 1.0 & 58.1 & 1.4 & 1.27 \\
2MASS J08173943--8243298 & 1.9 & 2.0 & 5.5 & 0.91 \\
2MASS J10284580--2830374 & 1.0 & 76.5 & 7.5 & 1.07 \\
2MASS J10285555+0050275 & 1.1 & 15.5 & 2.9 & 0.94 \\
2MASS J11020983--3430355 & 1.0 & 43.0 & 1.8 & 1.18 \\
2MASS J11210549--3845163 & 1.0 & 34.7 & 9.6 & 0.89 \\
2MASS J11393382--3040002 & 1.0 & 39.7 & 2.0 & 1.02 \\
2MASS J11395113--3159214 & 1.1 & 17.5 & 8.9 & 0.83 \\
2MASS J12072738--3247002 & 1.0 & 67.9 & 21.5 & 0.83 \\
2MASS J12073346--3932539 & 1.1 & 15.5 & 2.9 & 0.94 \\
2MASS J12153072--3948426 & 1.2 & 5.0 & 4.7 & 0.58 \\
2MASS J12265135--3316124 & 1.0 & 19.5 & 10.2 & 0.82 \\
2MASS J13215631--1052098 & 1.0 & 28.8 & 9.2 & 0.88 \\
2MASS J14112131--2119503 & 1.0 & 54.9 & 6.9 & 1.13 \\
2MASS J15385757--5742273 & 1.3 & 30.0 & 22.5 & 0.67 \\
2MASS J16334161--0933116 & 1.1 & 35.6 & 1.5 & 0.99 \\
FS2003 0979 & 1.0 & 25.9 & 1.5 & 0.95 \\
2MASS J18465255--6210366 & 1.3 & 18.7 & 1.3 & 1.13 \\
2MASS J19560294--3207186 & 1.0 & 58.4 & 1.1 & 1.33 \\
2MASS J19560438--3207376 & 1.0 & 65.6 & 4.0 & 0.67 \\
2MASS J20333759--2556521 & 1.0 & 27.2 & 1.3 & 1.15 \\
2MASS J20450949--3120266 & 1.0 & 54.4 & 3.3 & 0.54 \\
2MASS J21100535--1919573 & 1.0 & 55.8 & 1.4 & 0.93 \\
2MASS J21443012--6058389 & 1.3 & 24.8 & 1.1 & 1.20 \\
2MASS J22445794--3315015 & 1.2 & 0.0 & 5.0 & 0.94 \\
2MASS J22450004--3315258 & 1.2 & 0.0 & 5.0 & 0.94 \\
2MASS J23301341--2023271 & 1.0 & 54.3 & 0.9 & 2.10 \\
2MASS J23323085--1215513 & 1.5 & 0.0 & 4.9 & 1.33 \\
2MASS J23381743--4131037 & 1.1 & 38.6 & 2.5 & 0.65 \\ [1ex]
\hline
\end{tabular}
\label{conditions}
\end{center}
\end{table*}

\begin{table*}
\centering
\caption{$5 \sigma$ median contrast achieved within and beyond $0.5'$, and boundaries between the inner and outer regions for stars reduced with ADI.}
\begin{tabular}{c c c c}
\hline
Target & $\le 0.5'$ & $> 0.5'$ & Boundary \\ 
           &                 &               &   (as) \\ [0.5ex]
\hline
2MASS~J00243202--2522528 & 6.38 $\pm$ 0.06 & 8.34 $\pm$ 0.01 & 0.41\\ 
2MASS~J00251465--6130483 & 8.02 $\pm$ 0.06 & 10.04 $\pm$ 0.01 & 0.62\\
2MASS~J00255097--0957398 & 7.13 $\pm$ 0.03 & 8.14 $\pm$ 0.01 & 0.51\\ 
2MASS~J00452814--5137339 & 7.66 $\pm$ 0.08 & 9.52 $\pm$ 0.01 & 0.79\\ 
2MASS~J01033563--5515561 & 5.38 $\pm$ 0.19 & 7.91 $\pm$ 0.01 & \\ 
2MASS~J01071194--1935359 & 7.35 $\pm$ 0.19 & 10.03 $\pm$ 0.02 & 0.62\\ 
2MASS~J01231125--6921379 & 6.08 $\pm$ 0.06 & 6.32 $\pm$ 0.01 & \\ 
2MASS~J01365516--0647379 & 7.95 $\pm$ 0.06 & 8.84 $\pm$ 0.01 & 0.65\\ 
2MASS~J01521830--5950168 & 8.18 $\pm$ 0.04 & 9.32 $\pm$ 0.01 & 0.73\\ 
2MASS~J02224418--6022476 & 6.52 $\pm$ 0.10 & 9.12 $\pm$ 0.01 & \\ 
2MASS~J02365171--5203036 & 8.39 $\pm$ 0.06 & 9.84 $\pm$ 0.01 & 0.79\\ 
2MASS~J03350208+2342356 & 4.33 $\pm$ 0.08 & 4.58 $\pm$ 0.01 & \\ 
2MASS~J03363144--2619578 & 6.74 $\pm$ 0.08 & 7.69 $\pm$ 0.01 & \\ 
2MASS~J03472333--0158195 & 7.74 $\pm$ 0.06 & 10.19 $\pm$ 0.02 & 0.60\\ 
2MASS~J04141730--0906544 & 6.64 $\pm$ 0.09 & 7.54 $\pm$ 0.01 & \\ 
2MASS~J04373746--0229282 & 6.07$\pm$ 0.29 & 11.29 $\pm$ 0.02 & 0.38\\ 
2MASS~J04433761+0002051 & 6.08 $\pm$ 0.04 & 6.81 $\pm$ 0.01 & 0.38\\ 
2MASS~J04464970--6034109 & 8.44 $\pm$ 0.08 & 9.94 $\pm$ 0.01 & 0.79\\ 
2MASS~J04522441--1649219 & 9.03 $\pm$ 0.09 & 10.90 $\pm$ 0.01 & 0.95\\ 
2MASS~J04533054--5551318 & 6.10 $\pm$ 0.09 & 9.69 $\pm$ 0.03 & 1.00\\ 
2MASS~J04593483+0147007 & 6.93 $\pm$ 0.13 & 10.54 $\pm$ 0.02 & \\ 
2MASS~J05004714--5715255 & 8.18 $\pm$ 0.06 & 10.74 $\pm$ 0.02 & 0.68\\ 
2MASS~J05015881+0958587 & 7.73 $\pm$ 0.12 & 11.65 $\pm$ 0.02 & \\ 
2MASS~J05064991--2135091 & 7.67 $\pm$ 0.11 & 10.83 $\pm$ 0.02 & \\ 
2MASS~J05195412--0723359 & 6.30 $\pm$ 0.08 & 6.77 $\pm$ 0.01 & \\ 
2MASS~J05320450--0305291 & 8.03 $\pm$ 0.48 & 10.64 $\pm$ 0.01 & 0.84\\ 
2MASS~J06085283--2753583 & 5.18 $\pm$ 0.06 & 5.53 $\pm$ 0.01 & \\ 
2MASS~J06255610--6003273 & 8.56 $\pm$ 0.08 & 11.93 $\pm$ 0.02 & 0.57\\ 
2MASS~J07285117--3015527 & 8.11 $\pm$ 0.06 & 9.66 $\pm$ 0.01 & 0.60\\ 
2MASS~J07285137--3014490 & 7.66 $\pm$ 0.08 & 11.18 $\pm$ 0.02 & 1.06 \\ 
2MASS~J08173943--8243298 & 7.21 $\pm$ 0.12 & 10.67 $\pm$ 0.02 & \\ 
2MASS~J10284580--2830374 & 7.04 $\pm$ 0.04 & 8.03 $\pm$ 0.01 & 0.46\\ 
2MASS~J10285555+0050275 & 8.78 $\pm$ 0.06 & 11.80 $\pm$ 0.02 & 1.14\\ 
2MASS~J11020983--3430355 & 5.52 $\pm$ 0.03 & 5.94 $\pm$ 0.01 & 0.35\\ 
2MASS~J11210549--3845163 & 8.64 $\pm$ 0.05 & 9.66 $\pm$ 0.01 & 0.70\\ 
2MASS~J11393382--3040002 & 7.35 $\pm$ 0.06 & 8.70 $\pm$ 0.01 & 0.57\\ 
2MASS~J11395113--3159214 & 5.98 $\pm$ 0.08 & 6.49 $\pm$ 0.01 & \\ 
2MASS~J12072738--3247002 & 8.24 $\pm$ 0.07 & 10.02 $\pm$ 0.01 & 0.70\\ 
2MASS~J12073346--3932539 & 5.32 $\pm$ 0.08 & 6.14 $\pm$ 0.03 & \\ 
2MASS~J12153072--3948426 & 6.48 $\pm$ 0.09 & 9.43 $\pm$ 0.02 & \\ 
2MASS~J12265135--3316124 & 6.59 $\pm$ 0.05 & 7.63 $\pm$ 0.01 & 0.41\\ 
2MASS~J13215631--1052098 & 7.87 $\pm$ 0.09 & 9.06 $\pm$ 0.01 & 0.68\\ 
2MASS~J14112131--2119503 & 6.04 $\pm$ 0.07 & 6.54 $\pm$ 0.01 & 0.33\\ 
2MASS~J15385757--5742273 & 7.50 $\pm$ 0.04 & 10.44 $\pm$ 0.02 & 0.60\\ 
2MASS~J16334161--0933116 & 7.89 $\pm$ 0.06 & 9.77 $\pm$ 0.01 & 0.62\\ 
FS2003~0979 & 7.71 $\pm$ 0.06 & 9.04 $\pm$ 0.02 & 0.57\\
2MASS~J18465255--6210366 & 8.39 $\pm$ 0.04 & 9.36 $\pm$ 0.01 & 1.00\\ 
2MASS~J19560294--3207186 & 6.78 $\pm$ 0.42 & 9.49 $\pm$ 0.01 & 0.60\\ 
2MASS~J19560438--3207376 & 8.61 $\pm$ 0.06 & 9.61 $\pm$ 0.01 &0.70 \\ 
2MASS~J20333759--2556521 & 6.14 $\pm$ 0.07 & 7.94 $\pm$ 0.01 &0.57 \\ 
2MASS~J20450949--3120266 & 9.40 $\pm$ 0.08 & 12.41 $\pm$ 0.02 & 1.06\\  
2MASS~J21100535--1919573 & 7.55 $\pm$ 0.09 & 10.01 $\pm$ 0.02 &0.60 \\ 
2MASS~J21443012--6058389 & 8.58 $\pm$ 0.06 & 9.63 $\pm$ 0.01 &0.70\\ 
2MASS~J22445794--3315015 & 7.07 $\pm$ 0.10 & 10.37 $\pm$ 0.02 & \\ 
2MASS~J22450004--3315258 & 6.29 $\pm$ 0.13 & 8.96 $\pm$ 0.01 & \\ 
2MASS~J23301341--2023271 & 7.50 $\pm$ 0.08 & 10.44 $\pm$ 0.02 & 0.62\\ 
2MASS~J23323085--1215513 & 6.94 $\pm$ 0.12 & 10.02 $\pm$ 0.02 & \\ 
2MASS~J23381743--4131037 & 8.01 $\pm$ 0.09 & 11.63 $\pm$ 0.01 & 0.38\\ 
\end{tabular}
\label{contrasts}
\end{table*}

\section{Conclusion}
\label{Conclusion}
We conducted an adaptive optics survey for exoplanets orbiting nearby young low mass stars using $L^\prime$-band observations with NaCo at VLT. Two planetary mass companions were detected and previously reported  \citep{Delorme.2013,Chauvin.2004}, but no additional companions were detected. From our significant data set (54 targets), we derived the planetary mass companion (and substellar companion) frequency, defined here by the probability that a star hosts at least one planetary (substellar) mass companion. We used Bayesian statistics with an optimized conjugate prior, and we selected the sub-range in planet mass and semi-major axes where our data provides the best information, so that our statistics does not consider parameter ranges where our survey is not sensitive to companions. Our range of interest corresponds to planets more massive than 2~$M_\mathrm{Jup}$ and a semi-major axis between 8 and 400 AU. Using this range, we determined with a 68\% confidence that substellar companion frequency is $4.4^{+3.2}_{-1.3}$\%, and that planetary mass companion frequency is $2.3^{+2.9}_{-0.7}$\%. Considering all companions regardless of planet-to-star mass ratio, a Monte Carlo comparison between the substellar companion frequencies of the low-mass stars in the MASSIVE survey and a higher-mass A-F type star survey shows that there is a $\sim$~74\% probability that there are two distinct substellar companion populations, one orbiting around the low-mass stars and one orbiting around the higher-mass A-F type stars. We also found a $\sim$~75\% probability considering only low-mass ratios (Q<1\%). These results suggest that the frequency of imaged wide-orbit substellar companions is correlated with the stellar mass, in agreement with theoretical expectations \citep[see e.g.][]{Laughlin.2004}. For intermediate mass ratios (1\%<Q<5\%), the same comparison indicates a statistically significant similarity between the substellar companion populations around the low-mass star and high-mass star samples. We acknowledge that our results showing that substellar companion frequency could be correlated to stellar mass are still moderately significant. We therefore need more observations to confirm this important result.

  \begin{acknowledgements}
We thank the staff of ESO-VLT for their support at the telescope. We acknowledge support from the French National Research Agency (ANR) through the GuEPARD project grant ANR10-BLANC0504-01. We acknowledge financial support from ''Programme National de Physique Stellaire" (PNPS) of CNRS/INSU, France. We thank Didier Fraix-Burnet for our discussions on statistics. The research of J.E.S was supported by an appointment to the NASA Postdoctoral Program at NASA Ames Research Center, administered by Oak Ridge Associated Universities through a contract with NASA.
\end{acknowledgements}

\bibliographystyle{aa}
\bibliography{biball}

\begin{appendix} 
\section{ Statistical analysis}
\label{stat_analysis}
Our statistical formalism is based on works by \citet{Carson.2006}, \citet{Lafreniere.2007}, \citet{Vigan.2012}, and \citet{Rameau.2013}.
\\
Let us consider $N$, the total number of stars in a survey. From our detection limit analysis described in the text, the mean detection probability to find a companion of a given mass at a given semi-major axis, around a given target $j$ of a survey is $p_j$. $p_j$ is derived from the 2D detection limit maps. We denote $f$ the fraction of stars around which there is at least one planet, with a mass included in the interval [$m_{min},m_{max}$], and for separations inside the interval [$a_{min},a_{max}$]. $fp_j$ is the probability to detect a planetary mass companion (PMC) around the star $j$ given its mean detection probability $p_j$ and the fraction of stars $f$, and $1-fp_j$ is the probability not to find it. The detections and non-detections that are reported for a survey are denoted $\{d_j\}$: $d_j=1$ for stars around which we found a planet, and $0$ otherwise. The likelihood function of the data from which we can find $f$ is $L(\{d_j\}|f)$, which is the product of each Bernoulli event: this represents the probability model that gives \{$d_j$\} given the planet fraction $f$.

\begin{equation}
L(\{d_j\}|f)=\prod_{j=1}^N (1-fp_j)^{1-d_j} \times (fp_j)^{d_j}
\end{equation}

Bayes' theorem provides the probability density of $f$, the fraction of stars hosting at least one PMC given the observed data \{$d_j$\}: this probability density is the posterior distribution that represents the distribution of $f$ given the observed data \{$d_j$\}.

\begin{equation}
P(f|\{d_j\})=\frac{L(\{d_j\}|f)P(f)}{\int_0^1 L(\{d_j\}|f)P(f)df}
\end{equation}

Where $P(\{d_j\})=\int_0^1 L(\{d_j\}|f)P(f)df$ is called the marginalized likelihood.

Once $f$ is calculated, we determine the interval $[f_{min},f_{max}]$ of $f$ for a confidence level $CL$ as:
\begin{equation}
CL=\int_{f_{min}}^{f_{max}} P(f|\{d_j\})df
\end{equation}

An easier way to find $f_{min}$ and $f_{max}$ is to split the former equation into two others:

\begin{equation}
\frac{1-CL}{2}=\int_{f_{max}}^1 P(f|\{d_j\})df
\end{equation}

\begin{equation}
\frac{1-CL}{2}=\int_0^{f_{min}} P(f|\{d_j\})df
\end{equation}

In case of a null detection, the likelihood function is approximated using Poisson statistics: 

\begin{equation}
L(\{d_j\}|f)=\prod_{j=1}^N e^{-fp_j}
\end{equation}

Moreover, for such a null detection, $f_{min}=0$ and $f_{max}$ is given by:

\begin{equation}
f_{max}=\frac{-\ln(1-CL)}{N \left <p_j \right>}
\end{equation}

In the equation (.4), $P(f)$ is called the prior distribution. This prior is a distribution reporting any preexisting belief concerning the distribution of $f$. A first approach is to consider the prior distribution as a uniform distribution, $P(f)=1$. Another possibility is to define $P(f)$ as the natural conjugate prior of $P(f|\{d_j\})$: in that case, the prior and posterior distributions are from the same distribution family. Here, the posterior distribution is a Bernoulli distribution represented by a Beta function B: 

\begin{equation}
P(f)=\frac{(fp_j)^{\alpha -1}(1-fp_j)^{\beta -1}}{\mathrm{B(\alpha, \beta)}}
\end{equation}

$\alpha$ and $\beta$ are hyperparameters that can be fixed. We can then choose these hyperparameters such that $P(f)$ is mathematically identical to $L(\{d_j\}|f)$:

\begin{equation}
P(f)=L(\{d_j\}|f) \propto \prod_{j=1}^N (1-fp_j)^{1-d_j} \times (fp_j)^{d_j}
\end{equation}

Even if the mathematical forms of $P(f)$ and $L(\{d_j\}|f)$ are identical, $P(f)$ is a continuous function of $f$ given $d_j$, while $L(\{d_j\}|f)$ is a discrete function of \{$d_j$\} given $f$.

\section{Details on Monte Carlo methods using contrapositive logic.}
The following steps describe our Monte Carlo approach to determine the probability of a single giant planet population around the studied samples of low-mass and high-mass stars.
\\
0/ Before running our MC code, we calculate the Kolmogorov-Smironov statistic for each case (see Table\ref{MESS_bis}). \\
1/ First, we consider for each case both surveys and all the detections we have as if LMS and HMS belonged to the same survey. Thus, we consider here only one population of SCs. We then derive the frequency of SCs for each case (cf. Fig.\ref{proba_2surveys}).\\
2/ Second, using the merged distribution derived in 1/, we chose randomly the frequency $f$, for each case, that corresponds to the frequency of stars hosting at least one SC. We do it for 10000 iterations.\\
3/ Then, we attribute randomly to each star (for both surveys) zero or at least one substellar companion, using the frequency derived in 2/.\\
4/ If at least one substellar companion has been attributed to a star, we calculate if it has been detected using the mean probability to detect it around the considered star. Thus, after allocating each star to the survey it initially belonged to, each survey ends with n$_1$ and n$_2$ detections (respectively for the LMS and the HMS surveys).\\
5/ Once each original survey has been attributed n$_1$ and n$_2$ detections, we derive the frequency of SCs for the LMS and the HMS surveys, as we did previously. Then, we find two frequency ranges (for the LMS and the HMS surveys), that we will use in 6/ to know the extent to which both distributions, and thus populations, are different.\\
6/ We derive the Kolmogorov-Smironov statistic and its associated probability for all the iterations, for each case, using the frequency ranges found in 5/. We then count how many times the KS statistic is higher than the KS statistic derived for each case (cf. Table\ref{MESS_bis}): thus, we derive the probability $P$ that we can identify two distinct populations of SCs by chance when only one population would actually exist. \\
7/ We derive then the probability that there are two distinct populations of SCs when considering two distinct surveys at the beginning ($1-P$).

\section{2MASS~J08540240--3051366}
\label{J0854_par}
\begin{figure*}[h]
\includegraphics[width = 90mm]{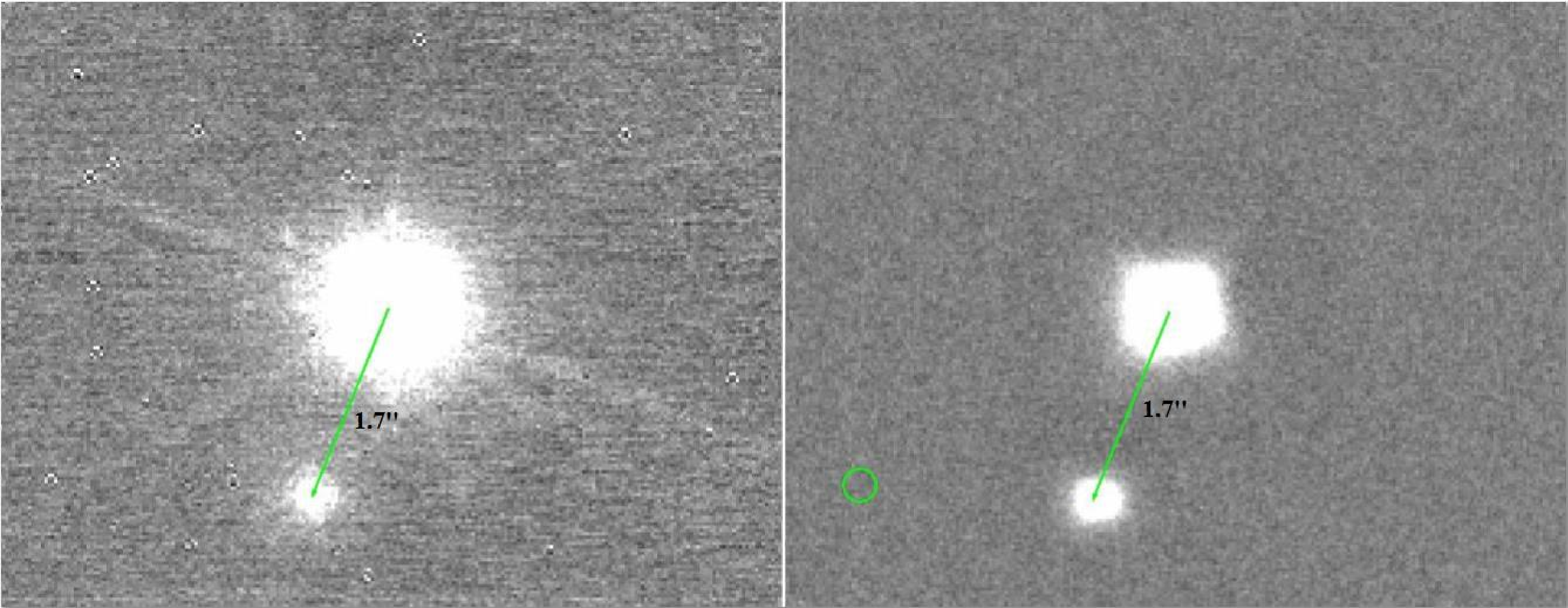}
\caption{\label{Fig.1} Left: J0854~A and B data from January 2006, using NaCo and the $K_S$ band. Right: J0854~A and B in November 2012, using NaCo in the $L^\prime$ band. The green arrow indicates the position of the companion in 2006, and the green circle displays the expected position of J0854~(AB)~C if it were a background field star. East is left and North is up.}
\label{J0854}
\end{figure*}

This target was identified as a probable young star by FS2003 \citep{Fuhrmeister.2003}. We used the BANYAN~II Bayesian analysis tool \citep{Gagne.2015} to assess whether it is a probable member of a young moving group. Using the proper motion that we measured from the 2MASS and AllWISE astrometry epochs ($\mu_{\alpha}\cos\delta = -275.3 \pm 5.5$\masyr; $\mu_{\delta} = -14.7 \pm 6.1$\,\masyr) with its sky position and 2MASS and AllWISE photometry, we obtained a 89\% membership probability to the $\beta$~Pictoris moving group (BPMG), associated with a 38\% probability for it to be an unresolved binary star. The statistical distance and radial velocity (RV) associated with a membership to $\beta$~Pictoris are $10.9 \pm 1.2$\,pc and $15.6 \pm 1.7$\,km\,s$^{\rm -1}$, respectively.
\\
We observed the star with NaCo in November. 2012 (090.C-0698(A)) and obtained additional NaCo observations from the ESO archive observed in January 2006. After data reduction and analysis, we identified a comoving companion orbiting this target (see Figure~\ref{J0854}). We measured absolute magnitudes of $L^\prime=9.50 \pm 0.05$, $K_S=10.03 \pm 0.01$, $H=10.44 \pm 0.01$, and $J=11.05 \pm 0.03$ for the companion, which orbits at a separation of $\approx 26$\,AU from its host star ($\approx 1.7$\,as with a position angle of $158.7 \pm 0.3^\circ$ from the $L^\prime$-band data, and $156.2 \pm 0.3^\circ$ from the $K_S$-band data). The NIR colours of the companion are both consistent with a field-age or a young object, hence the photometric NaCo data cannot further constrain its age.
\\
We subsequently obtained NIR spectra for the host and companion using FLAMINGOS II (GS-2013B-Q-79, PI: J.~Gagn\'e) and SINFONI (292.C-5036A, PI: J.~Lannier), which allowed us to assign a spectral type of M4 to the host star. This NIR spectrum did not allow us to place constraints on the age of the host star. Contrary to the BANYAN~II prediction according to which the companion would be a probable young object near the planetary-mass boundary, its NIR spectrum is similar to M8 field stars, with no apparent signs of low gravity. The spectral types mentioned above were obtained using the visual classification method described by \citet[][BANYAN~VII]{Gagne.2015}, and are both associated with an uncertainty of 0.5 subtypes.
\\
In addition to the observations mentioned above, we obtained three high-resolution optical spectra ($R \simeq 115000$; 3800--6900\,\AA) of the host star with the HARPS spectrograph located at the 3.6\,m ESO telescope at La Silla observatory (LP192.C--0224, PI: Lagrange), in order to measure its systemic RV and improve our BANYAN~II membership probability assessment. We used the automated HARPS Data Reduction System to obtain a first RV measurement estimate, and identified high-amplitude variations of the order of several tens of km\,s$^{\rm -1}$ in the RV signal. The cross-correlation functions (CCF) associated with these spectra show two peaks, indicative of a double-lined spectroscopic binary (SB2). The $H_{\alpha}$ emission line also displays a double-peak shape.
\\
We found that the two peaks of the CCF display significantly different depths, which allowed us to distinguish between the two components of the SB2 host star and to retrieve individual RV measurements. These values are summarized Tab.~\ref{RV}. For each component, the measurement RV corresponds to the sum of the absolute RV of the binary $V_{\rm~\gamma}$ and that of the relative RV of the component with respect to $V_{\rm~\gamma}$. The relative RV is proportional to the phase of the binary system motion; we can thus obtain measurement of $V_{\gamma}$ using each spectrum. The values that we obtain are summarized in Table \ref{RV}. We take the mean value of these three systemic RV measurement to obtain $V_{\gamma} = 45 \pm 1$\,km\,s$^{\rm -1}$, where the error is an empirical estimate.
\\
These new observations strongly reject any possible membership to BPMG, both in terms of age and kinematics. The lack of low-gravity features in the NIR spectrum of the M8-type companion 2MASS~J0854~(AB)~C constraints the age of the system at $\gtrsim 250$\,Myr \citep{Allers.2013}, whereas the systemic RV measurements reduces the Bayesian probability of a membership to any young moving group to zero using BANYAN~II (i.e., TW~Hydrae, BPMG, Tucana-Horologium, Carina, Columba, Argus, and AB~Doradus). We have used a new Bayesian analysis tool that includes more moving groups and associations (BANYAN--M; J.~Gagn\'e et al., in preparation, see also A.~Riedel et al., submitted to the Astronomical Journal) to further reject any possible membership to the Hercules-Lyrae, Carina-Near, Ursa Major, 32~Ori, Alessi~13, $\varepsilon$~Cha, $\eta$~Cha, Coma~Berenices associations and moving groups, as well as to the Pleiades association.
\\
The binary nature of the host star 2MASS~J0854~(AB) reconciles the spectral type of the companion that seemed too early with respect to its contrast ratio to the host star. This system is thus in all likelihood a field-age SB2 binary M4-type host star with a wide-separation M8-type stellar companion that display a proper motion similar to BPMG members by pure chance. Such interlopers are expected in BANYAN~II candidate members \citep[e.g., see][BANYAN~II]{Gagne.2014}, which outlines the necessity of obtaining full kinematics and youth indications before declaring an object as a new bona fide member of a young moving group. We have thus excluded the 2MASS~J0854~(AB)~C system from our survey statistic analysis described in Section~\ref{Frequency}.

\begin{table*}[h]
\centering
\caption{Radial velocities ($km.s^{-1}$) obtained with HARPS}
\begin{tabular}{c c c c}
\hline\hline
Observation dates & Primary RV & Secondary RV & $V_{\rm~\gamma}$\\ [0.5ex]
\hline
 2013-12-05 & 82.9 & 4.3 & 44.6 \\
 2014-02-16 (1) & 28.4 & 62.9 & 45.2 \\
 2014-02-16 (2) & 2.7 & 89.3 & 44.9 \\
\end{tabular}
\label{RV}
\end{table*}

\end{appendix}


\end{document}